\pdfoutput=1
\documentclass[%
reprint,
superscriptaddress,
prstab,
floatfix,
]{revtex4-1}

\usepackage{amssymb,amsmath}
\usepackage{graphicx}
\usepackage[tight,footnotesize]{subfigure}
\usepackage{dcolumn}
\usepackage{bm}
\usepackage[utf8]{inputenc}
\usepackage{gensymb}
\usepackage{units}

\def\mathclap#1{\text{\hbox to 0pt{\hss$\mathsurround=0pt#1$\hss}}} 
\newcommand{\msub}[1]{_{\text{#1}}}
\newcommand{\musb}[1]{_{\text{#1}}}

\newcommand{\refF}[1]{Fig.\,\ref{#1}}

\newcommand{\refE}[1]{Eq.\,\ref{#1}}

\DeclareFontFamily{U}{euc}{}
\DeclareFontShape{U}{euc}{m}{n}{<-6>eurm5<6-8>eurm7<8->eurm10}{}%
\DeclareSymbolFont{AMSc}{U}{euc}{m}{n} 
\DeclareMathSymbol{\umu}{\mathord}{AMSc}{"16}

\begin{document}
\preprint{APS/123-QED}

\title{Field of first magnetic flux entry and pinning strength of superconductors for RF application measured with muon spin rotation }

\author{T. Junginger} 
\email{Tobias.Junginger@lancaster.ac.uk}
\affiliation{Engineering Department, Lancaster University, UK}
\affiliation{Cockcroft Institute, Warrington, UK}
\author{S.H. Abidi}
\affiliation{University of Toronto, Toronto, Canada}
\author{R. Astley}
\affiliation{University of Waterloo, Canada}
\author{T. Buck}
\affiliation{University of British Columbia, Vancouver, Canada}
\author{M.H. Dehn}
\affiliation{University of British Columbia, Vancouver, Canada}
\author{S. Gheidi}
\affiliation{University of Toronto, Toronto, Canada}
\author{R. Kiefl}
\affiliation{University of British Columbia, Vancouver, Canada}
\affiliation{TRIUMF Canada's National Laboratory for Particle and Nuclear Physics, Vancouver}
\author{P. Kolb}
\affiliation{Brookhaven National Laboratory, USA}
\author{D. Storey}
\affiliation{University of Victoria, Canada}
\affiliation{TRIUMF Canada's National Laboratory for Particle and Nuclear Physics, Vancouver}
\author{E. Thoeng}
\affiliation{University of British Columbia, Vancouver, Canada}
\affiliation{TRIUMF Canada's National Laboratory for Particle and Nuclear Physics, Vancouver}
\author{W. Wasserman}
\affiliation{University of British Columbia, Vancouver, Canada}
\author{R.E. Laxdal}
\affiliation{TRIUMF Canada's National Laboratory for Particle and Nuclear Physics, Vancouver}

\date{\today}

\begin{abstract}
The performance of superconducting radiofrequency (SRF) cavities used for particle accelerators depends on two characteristic material parameters: field of first flux entry $H\msub{entry}$  and pinning strength. The former sets the limit for the maximum achievable accelerating gradient, while the latter determines how efficiently flux can be expelled related to the maximum achievable quality factor. In this paper, a method based on muon spin rotation ($\mu$SR) is developed to probe these parameters on samples. It combines measurements from two different spectrometers, one being specifically built for these studies and samples of different geometries. It is found that annealing at \unit[1400]{$\degree$C} virtually eliminates all pinning. Such an annealed substrate is ideally suited to measure $H\msub{entry}$ of layered superconductors, which might enable accelerating gradients beyond bulk niobium technology. 

\end{abstract}

\maketitle

\section{Introduction}

\begin{figure}[tbh]
   \centering
	 \includegraphics[width=0.9\columnwidth]{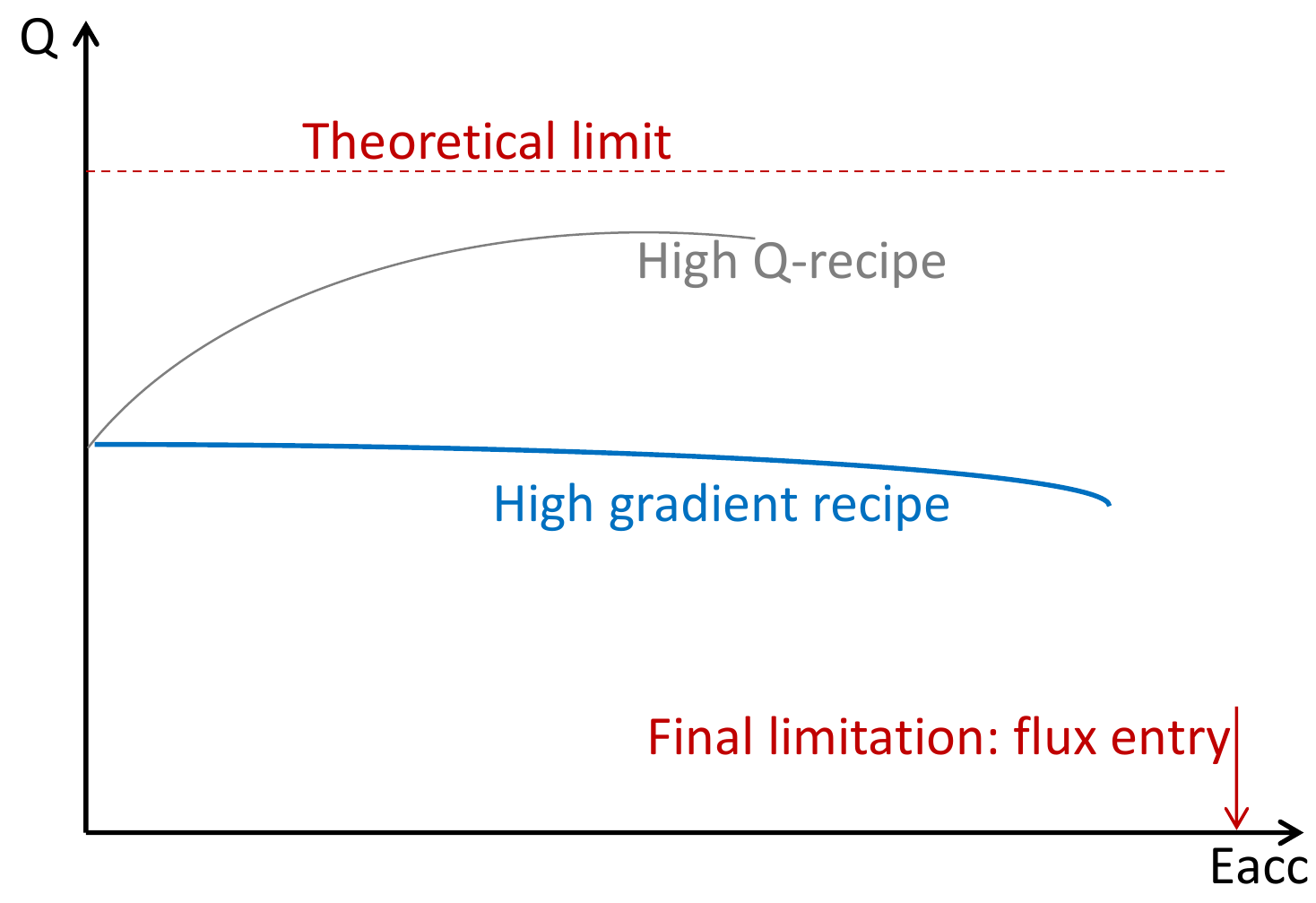}
   \caption{Generic plot of the quality factor Q as a function of the accelerating gradient.}
	\label{fig:QvsE}
\end{figure}

SRF cavities have been used to increase the energy of charged particles for more than 50 years \cite{Padamsee:1116813}. The material of choice is niobium, the element with the highest critical temperature and critical fields. The performance of these cavities is usually expressed in a plot of the quality factor as a function of the accelerating gradient, see Fig. \ref{fig:QvsE}. The maximum achievable value for both quantities is related to the intrinsic material properties but also to the surface preparation. Depending on application, different recipes consisting of baking - under vacuum or in a gas atmosphere - and chemical treatments are applied. 

The maximum accelerating gradient can be limited by several mechanisms including field emission, quench, and a strong decrease of the quality factor with accelerating gradient at high fields (Q-drop); in general the critical field of the material is not reached. The most  common `recipe' for preparing cavities is the so-called `ILC' or `High Gradient Recipe' where the cavities receive a deep etch of \unit[120]{$\mu$m} by either electropolishing (EP) or buffered chemical polishing (BCP), followed by baking at \unit[800]{{\degree}C} for \unit[4]{hours} to degas hydrogen and a final flash EP of \unit[5-10]{$\mu$m}. A final low temperature bake-out at \unit[120]{{\degree}C} for \unit[24-48]{hours} in vacuum is applied to increase the peak field performance \cite{Padamsee:1180071}. Cavities prepared by this `high gradient recipe' are usually limited by field emission or quench. These limiting mechanisms are specific to RF fields and related to the cleanliness of the surface and contaminants. They are not fundamental limitations of the material itself. It would be beneficial to characterize materials in terms of $H\msub{entry}$ as a function of surface and bulk treatments using DC methods without having to build  an entire cavity. One potential method would be magnetometry. However, interpretations of results obtained by this technique are often ambiguous due to geometrical effects and pinning. Muon spin rotation and relaxation ($\mu$SR) is an alternative method that can be used to directly monitor the magnetic  field inside the sample. It is a local probe which in principle can detect the  field  at specific locations in the sample. As such  it provides information  which is complementary to bulk methods such as magnetometry.

Recently, to reach high quality factors, a treatment procedure has been established baking cavities at \unit[800]{$\degree$C} and injecting nitrogen gas at the end of this treatment. Consequently, cavities receive a light chemical etch to remove the outermost layer \cite{grassellino2013nitrogen}. This `high Q-recipe' limits the accelerating gradients to lower values than the `high-gradient recipe' but enables quality factors close to the theoretical limit set by losses from thermally activated quasiparticles. These are fundamental to superconductors operated under RF fields above \unit[0]{K}. To achieve highest quality factors especially with such cavities it is necessary to avoid trapping of external magnetic flux. Generally magnetic shielding is applied to reduce the earth's magnetic field to a small fraction, but for ultimate performance, expulsion of the residual flux is necessary. Flux expulsion depends on the cooling dynamics around the critical temperature of the material $T\msub{c}$ and its pinning strength. The $\mu$SR technique allows measurement of the magnetic flux inside a sample. By choosing an appropriate sample and field configuration, it  enables measurement of the pinning strength of test samples.       

The first application of $\mu$SR to SRF materials has been reported in 2013. Using the TRIUMF surface muon beam, Grassellino et al. \cite{Grassellino2013} characterized samples cut out from cavities using the LAMPF spectrometer. These studies used a geometry that allowed comparison of the pinning strength of the different samples. In this paper we present complementary studies that aim to reveal the field of first flux entry $H\msub{entry}$. For this purpose samples of ellipsoidal geometry have been produced. In the experiments reported in \cite{Grassellino2013}, the magnetic field was applied perpendicular to the sample surface, unlike in accelerating cavities. To resemble the field geometry of SRF cavities, a spectrometer that allows the application of fields of up to \unit[300]{mT} parallel to the sample surface was built. The combination of the different sample shapes and field geometries now allows the determination of the field of first flux entry and the pinning strength of the same samples as a function of surface and bulk treatments. 

\section{Experimental setup and technique}
Muon spin rotation is a powerful condensed matter technique with many applications in magnetism and superconductivity. For example it can be used to understand superconductors in terms of their magnetic-phase diagram and penetration depth, as well as to characterize impurities based on muon diffusion. In the early 1970s, new high-intensity, intermediate-energy accelerators were built at PSI (Paul Scherrer Institute), TRIUMF (TRI-University Meson Facility), and LAMPF (Los Alamos Meson Physics Facility). These new `meson factories' produced pions (and therefore muons) at a rate several orders of magnitude more than previous sources - and in doing so, ushered in a new era in the techniques and applications of $\mu$SR.
 
\begin{figure}[tbh]
   \centering
	 \includegraphics[width=0.9\columnwidth]{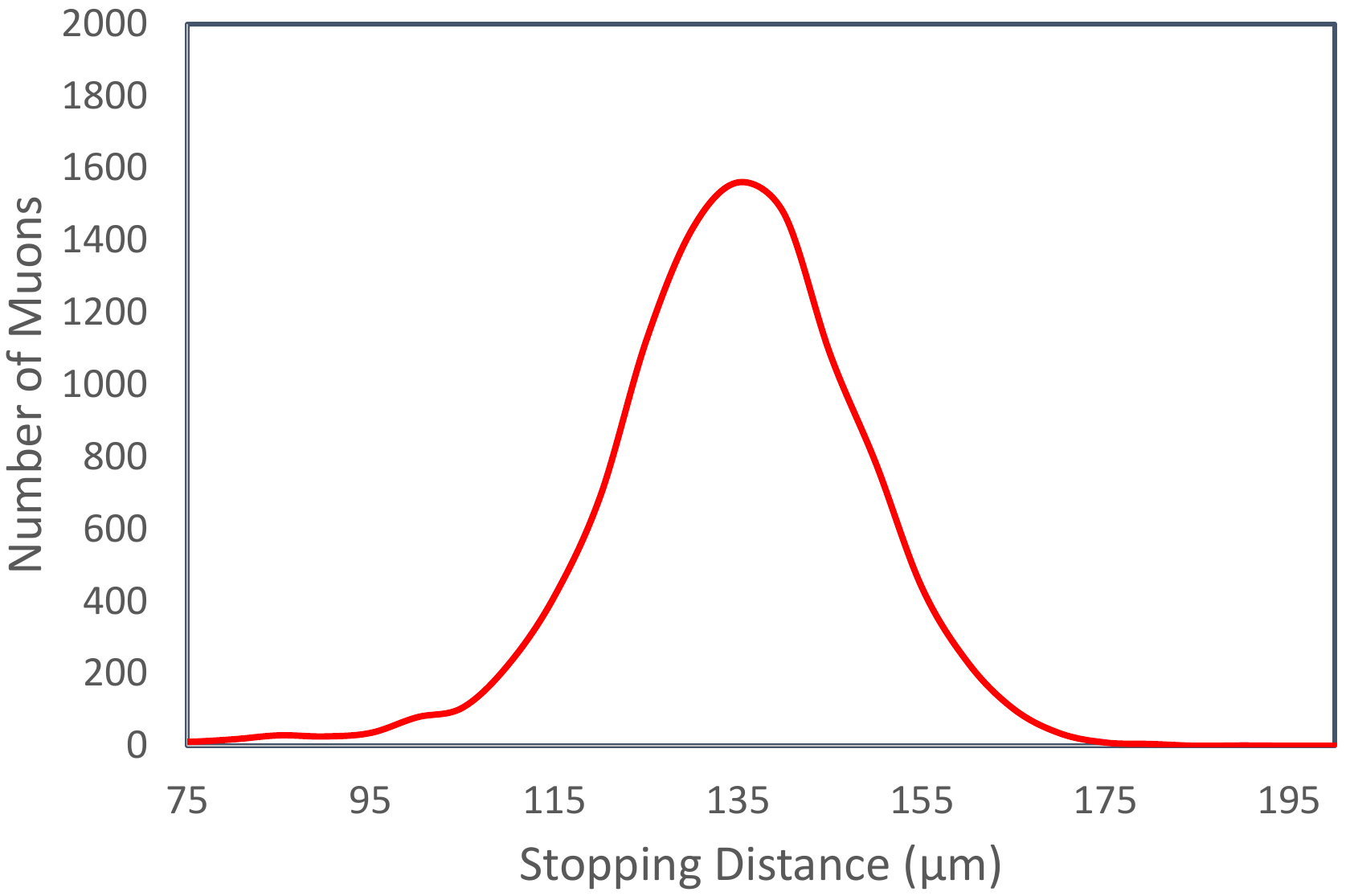}
   \caption{Muon stopping distance in Nb as simulated by TRIM. The simulation takes into account all obstacles the muons encounter in their path such as beamline windows and scintillators.}
	\label{fig:muoninNB1}
\end{figure}

For the experiments presented here, surface muons are emitted from a production target \unit[100]{\%} spin polarized with momentum and energy of \unit[29.8]{MeV/c} and \unit[4.1]{MeV}+/-6\% respectively. They are implanted one at a time into the sample. These muons have an average stopping distance of \unit[130]{$\mu$m} in niobium, as simulated by TRIM \cite{ziegler2010srim}, see Fig. \ref{fig:muoninNB1}. 

When the muon decays (half life=\unit[2.197]{$\mu$s}), it emits a fast positron preferentially along the direction of its spin at the time of the decay. By detecting the rate of emitted positrons as a function of time with two detectors placed symmetrically around the sample, here `up' and `down', the time evolution of the spin precession of the muon and therefore the magnetic field properties experienced by the muon can be inferred from the time dependent asymmetry in the positron decay 
\begin{equation}
	Asy(t)=\frac{N\msub{U}(t)-\alpha N\msub{D}(t)}{N\msub{U}(t)+\alpha N\msub{D}(t)}=A\cdot P(t).
	\label{eq:Alpha}
\end{equation}
Here, $N\msub{U}(t)$ is the number of counts in the `up' detector and $N\msub{D}(t)$ is the number of counts in the `down' detector. The parameter $\alpha$ is added to account for detector efficiencies and to remove any bias caused by uneven solid angles. In the case where the detector efficiencies are identical, $\alpha$ assumes a value of 1, $A$ is the initial asymmetry, while the depolarization function $P(t)$ signifies the change of asymmetry with time.

The aim of this experiment is to measure the fraction of the surface area probed by the muon beam which is in a field free Meissner state. Samples are placed in a cryostat surrounded by field inducing coils. For field penetration measurements, samples are cooled to below $T\msub{c}$ (\unit[2.5]{K} is common) in zero field and then a static magnetic field is applied perpendicular to the initial spin polarization to probe if field has penetrated the sample. Specifically, the polarization signal gives information on the volume fraction of the host material sampled by the muon that does not contain magnetic field. This signal can be used to characterize the superconducting state, particularly the transition from Meissner to mixed state. 

\subsection{Measurements in strong parallel fields}
\begin{figure}
  \centering
  \subfigure[\label{a}]{\includegraphics[width=\columnwidth,height=4cm]{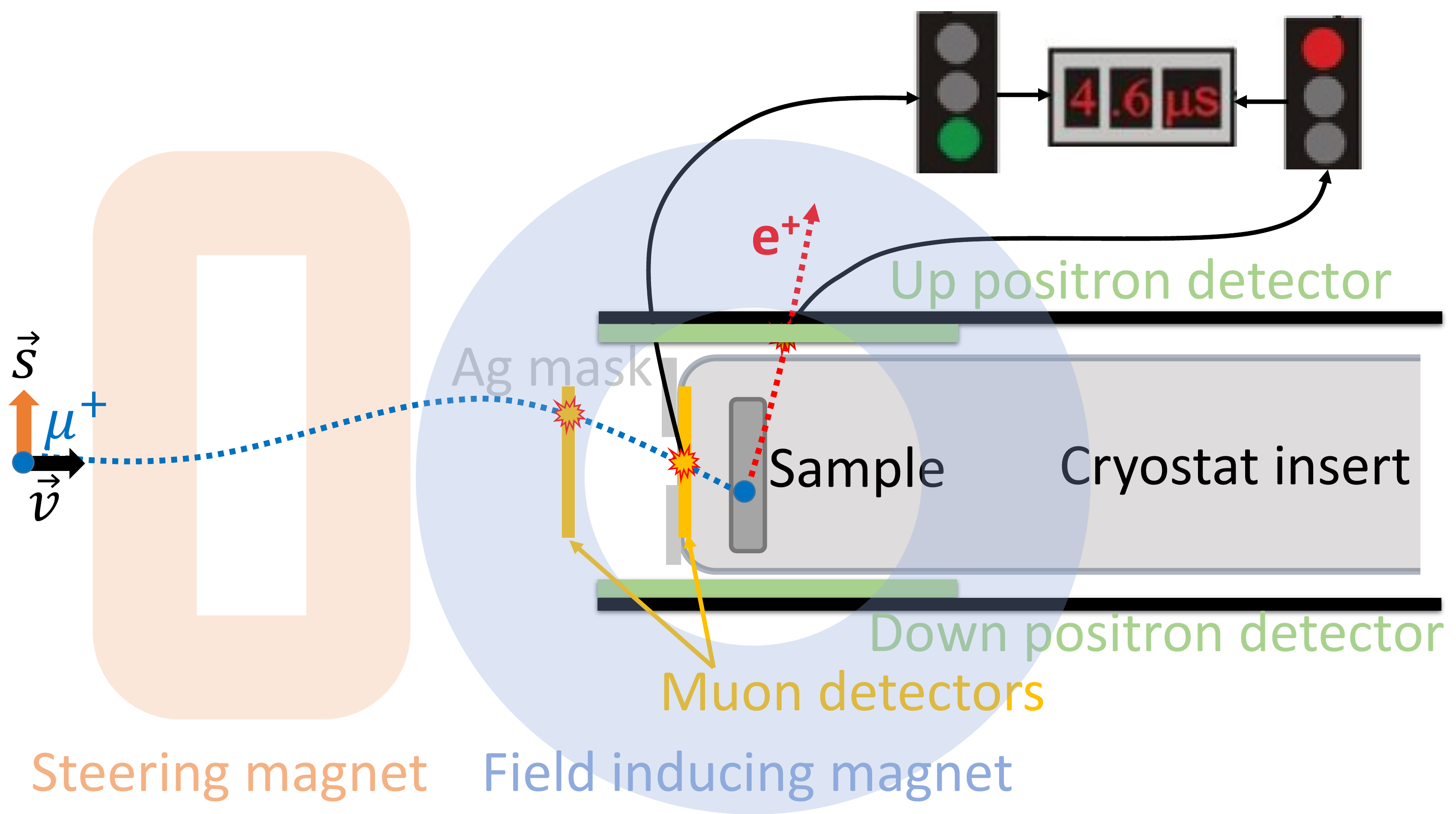}}\quad
	\subfigure[\label{b}]{\includegraphics[width=\columnwidth]{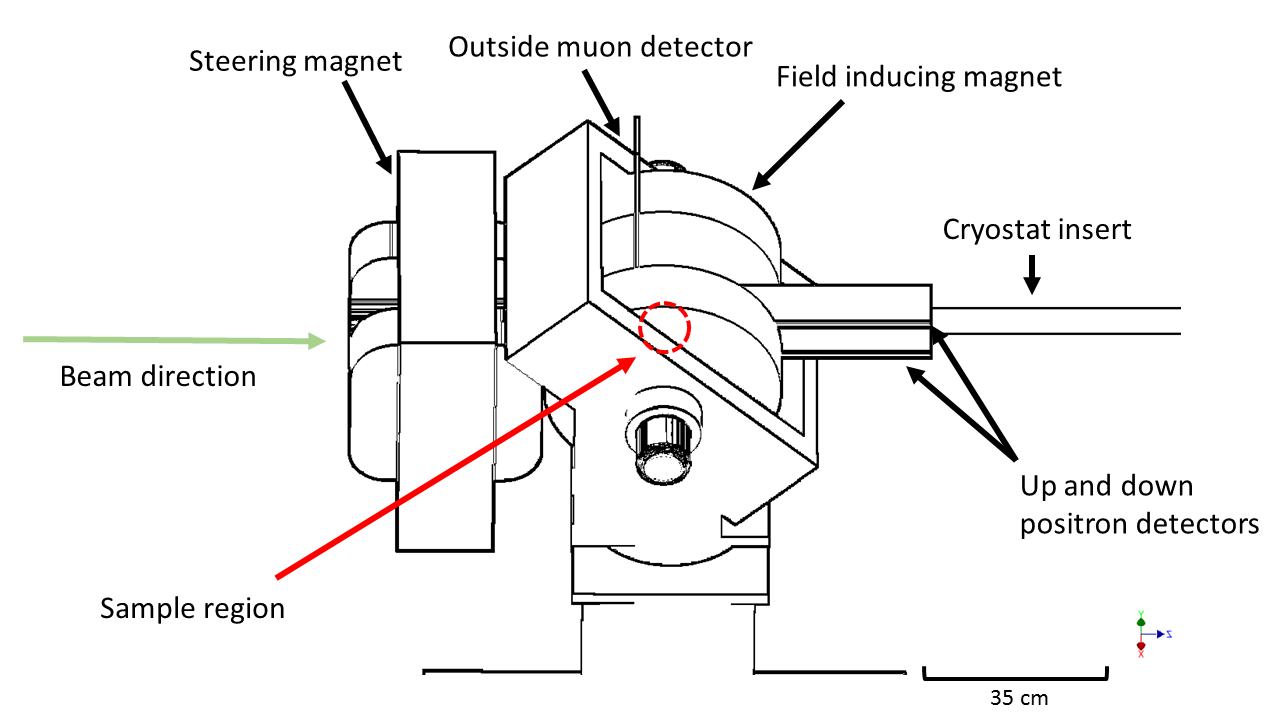}}
  
  \caption{a) Schematic displaying the components of the HPF spectrometer and the beam trajectory b) 3D render of the spectrometer.}
\label{fig:HodgePodge}
\end{figure}

In order to allow for measurements in parallel magnetic fields of up to \unit[300]{mT}, resembling the field geometry of SRF cavities, a dedicated spectrometer named High Parallel Field apparatus (HPF) has been added to the TRIUMF $\mu$SR M20 facility \cite{gheidi2015muon}. Due to the presence of the Lorentz force experienced by the muons in parallel geometry, the field that is used to probe the sample also bends the muon trajectory. Therefore, an upstream steering magnet is used to pre-steer off-axis and the applied field at the sample bends the particles back to the sample, see \refF{fig:HodgePodge}. Passing an initial muon counter (scintillator) an electronic clock is started, a silver mask with an \unit[8]{mm} diameter hole in the center is used to restrict the measured muons to the center of the sample. A second muon counter behind the silver mask probes whether the muon went through the hole or was stopped by the mask. The spin-polarized muons are implanted into the sample, and quickly stop at interstitial sites in the bulk. The muon intensity and count rate are such that the muons are counted one by one with the positron decay stopping the clock for each muon. 



\subsection{Polarization functions}

If no magnetic field has entered the sample the depolarization of the muons is caused by the internal dipolar fields. Since these fields are randomly distributed, each muon will sense a different field orientation resulting in a quick loss of polarization. In the case of a static random field distribution, e.g. nuclear dipole fields, and the absence of muon diffusion, the muon spin polarization is given by the static zero field Gaussian Kubo-Tuyabe function \cite{Hayano}
\begin{equation}
 P_{\rm ZF}^{\rm stat.}(t) = \frac{1}{3} + \frac{2}{3} \left[ 1 - (\sigma t)^2 \right] \exp\left[-\frac{1}{2} (\sigma t)^2\right],
\label{eq:StatGssKT}
\end{equation}
where $\sigma$ is the width of the dipolar field distribution. The function is characterized by an initial Gaussian shape and assumes 1/3 for long times. The initial part is explained by the Gaussian distributed nuclear magnetic dipolar fields from neighboring Nb nuclear spins that influence the muon spins. The relaxation to 1/3 of the initial value is due to the component of local fields along initial direction of polarization, i.e. 1/3 of the muons are polarized along the axis of initial polarization \cite{Hayano}. 
Equation \ref{eq:StatGssKT} is only applicable to muons being static after initial trapping in static fields. Internal field dynamics, resulting either from the muon hopping from site to site or from fluctuations of the internal fields themselves, can be accounted for by using the strong collision approximation. This model assumes that the local field changes its direction at a time $t$ according to a probability distribution 
\begin{equation}
p(t)=\exp(-\nu t),
\end{equation}
with the hop rate $\nu$. In the strong collision model, the field after each `collision' assumes a random value from the internal distribution, entirely uncorrelated with the field before the
collision. The resulting expression is the dynamic Kubo-Toyabe (dynKT) depolarization function \cite{Hayano} 
\begin{eqnarray}\label{eq:strong-collision-model}
  P_{\rm ZF}^{\rm dyn.}(t) &=& P_{\rm ZF}^{\rm stat.}(t) \exp(-\nu t) +  \\ \nonumber 
       & & +\, \nu \int_0^t dt^\prime \left\{ P(t-t^\prime) P_{\rm ZF}^{\rm stat.}(t^\prime) \exp(-\nu t^\prime) \right\},
\end{eqnarray}

For large values of $\nu$ $P(T)$ will assume an exponential decay shape and the recovery to 1/3 is completely suppressed. An overview of depolarization functions commonly used in muon spin rotation experiments can be found in \cite{yaouanc2011muon}.  
\begin{figure}[htbp]
    \centering
    \includegraphics[width=\columnwidth]{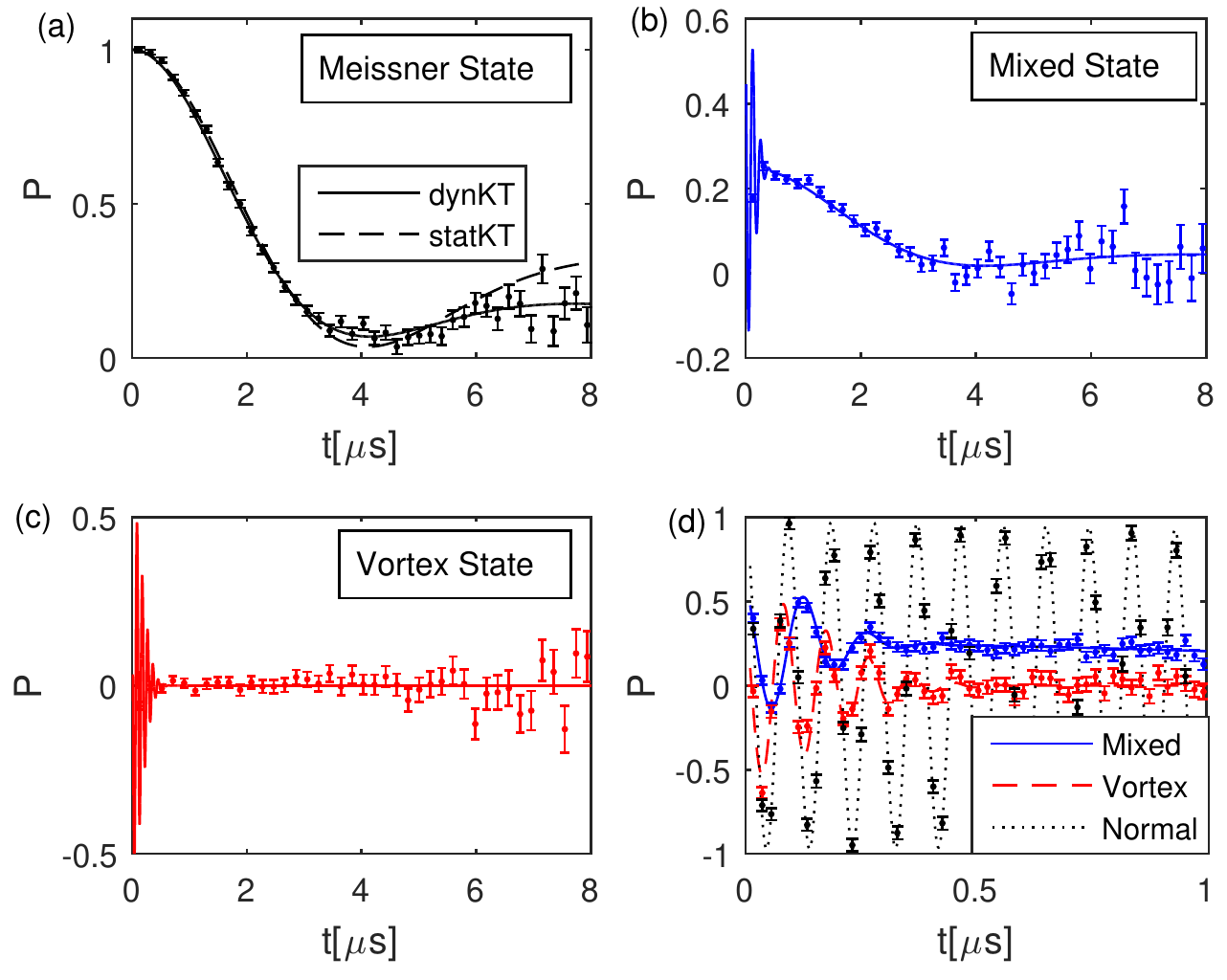}
    \caption{Polarization function in different states. (a) In the Meissner state the depolarization is caused by randomly orientated nuclear dipole fields resulting in the characteristic dynamic Kubo-Tuyabe polarization function. For comparision a fit to the static Kubo-Tuyabe polarization function is plotted as well. This fit does not give a good representation for longer times. (b) In the mixed state the Kubo-Tuyabe polarization function is combined with a fast decaying oscillating function. (c) In the vortex state all muons sense the external field and there is no signature of the Kubo-Tuyabe polarization function left. Here the strong damping is caused by the non uniformity of the vortex field structure. (d) In the normal state all muons probe the same field yielding weaker damping. Note the different time scale in this subplot.}
    \label{fig:States}
\end{figure}

In the case of RRR niobium (RRR$>$300), the muon is substantially diffusing in the material and the dynamic Gaussian-Kubo-Tuyabe function Eq.\,\ref{eq:strong-collision-model} is applicable, see \refF{fig:States}(a). When completely in the Meissner state, there is no field in the sample and $f_0$ is maximized. As the applied field is increased, flux begins to penetrate the sample. As a result of its influence on the precession frequency of the muons, the amplitude of the Gaussian portion of the Kubo-Toyabe function decreases. The total polarization function will become a sum of two terms. The first one is the dynamic Kubo-Tuyabe function with a reduced $f_0$. The second term is a damped oscillating function yielding the complete polarization function as displayed in \refF{fig:States}(b):
\begin{eqnarray}
\label{eq:Asy}
P(t)&=&f_0\cdot P_{\rm ZF}^{\rm dyn.}(t) + \\ \nonumber
& & f_1\cdot\exp{\left( -\frac{1}{2}\Delta^2t^2\right) }\cdot \cos{\left( \omega t + \frac{\pi\phi}{180}\right) }
\end{eqnarray}
with 
\begin{equation}
\omega=2 \pi \gamma_\mu H\msub{int},
\label{eq:omega}
\end{equation}
where $\gamma_\mu$=13.55KHz/G is the gyromagnetic ratio of the muon and $\phi$ a phase which can depend on the external field. The internal field $H\msub{int}$ has to be interpreted as the most probable internal field seen by the muons. In the normal state $H\msub{int}$=$H\msub{a}$ holds. For the mixed and the vortex state this is not the case. Here, $H\msub{int}$ depends on the structure of the vortex shape. For a detailed description how the polarization function depends on the vortex shape refer to Ref. \cite{Sonier2007}. The value of $f_0$ compared to its initial low field value is a measure of the volume fraction being in the field free Meissner state. Upon transitioning to the mixed state, $P(t)$ assumes the form of a heavily damped oscillation (\refF{fig:States}(c)).
Now the muons precess with varying frequencies that depend on their distance from the intruding vortices. As the field strength increases further, $f_0$ will assume 0 when the whole area probed by the muons is in the vortex state. The damping of the oscillation will eventually become much weaker signifying that sample is in the normal state. In this state, the polarization implies that the muons are precessing largely with the same frequency since magnetic flux affects all sites almost uniformly. It is the existence of nuclear dipolar fields which adds slight damping to the signal. 

\subsection{Normal state calibration}
We define the field of first flux entry when $f_0$ assumes a value significantly lower compared to its value at zero field. In the case of a pin free sample with no geometric edge boundary, that will happen suddenly in a sharp transition. Geometry and impurities can delay the flux penetration as mentioned above. Additionally, $f_0$ will also decrease as a function of field in the Meissner state since the muon will also precess in the external field outside of the sample before implantation. This can be accounted for by measuring the phase $\phi$ above the critical temperature $T\musb{c}$ as a function of the applied field $H\msub{a}$. The relation $\phi(H\msub{a})$ is subsequently used to correct the measured values of $a\msub{0$\mid$measured}$ to physical meaningful values

\begin{equation}
f_0(H\msub{a})=\frac{f\msub{0$\mid$measured}}{\cos{\phi(H\msub{a})}}.
\end{equation}

\begin{figure}[htbp]
    \centering
    \includegraphics[width=\columnwidth]{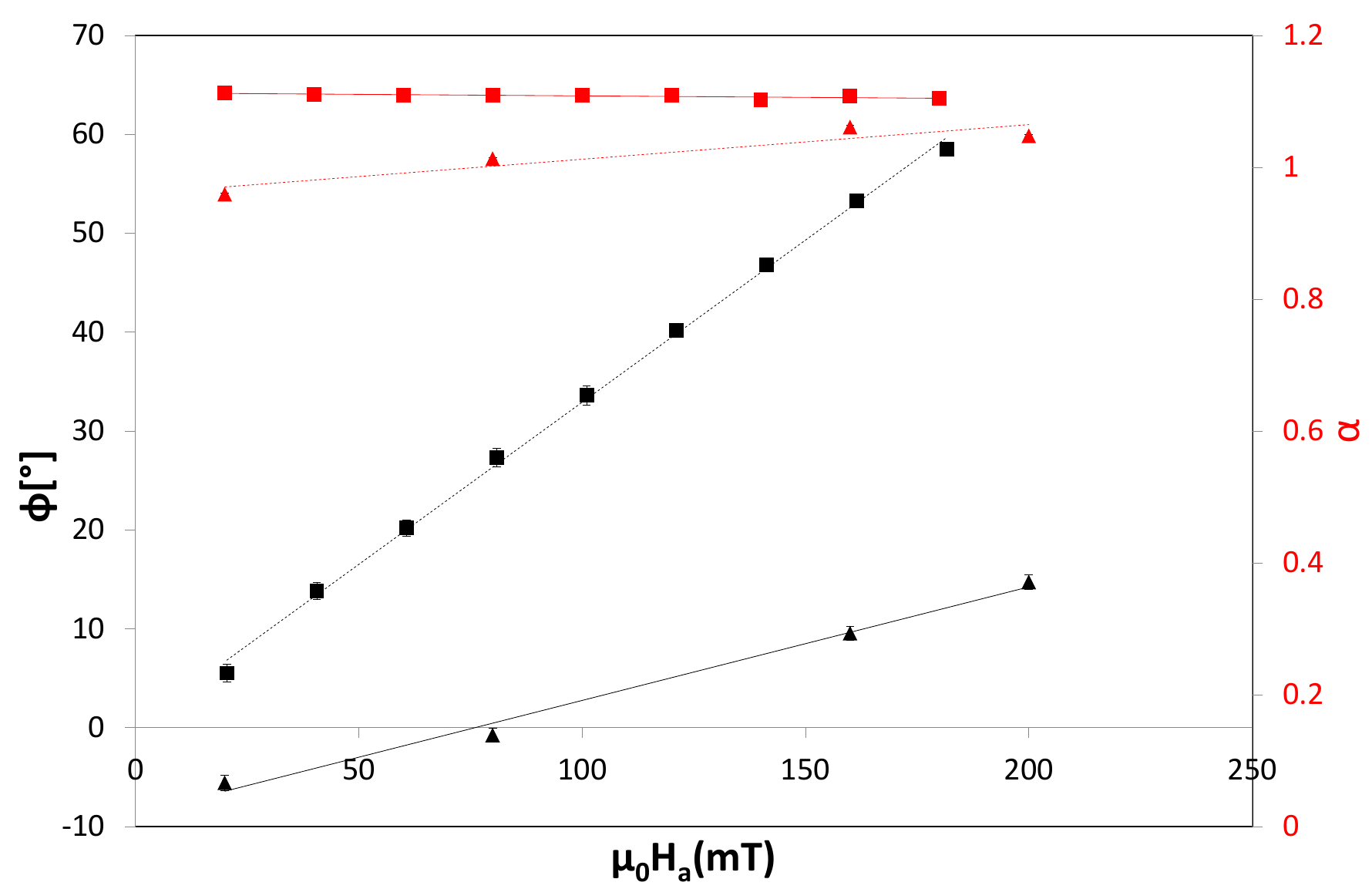}
    \caption{Phase $\phi$ and $\alpha$ as a function of applied field $H\msub{a}$ above $T\msub{c}$. Diamonds/Squares are for the HPF/LAMPF spectrometer.}
    \label{fig:Callibration}
\end{figure}

Since the rotation of the muon is proportional to the magnetic field strength, a linear relation between $\phi(H\msub{a})$ is expected and could be experimentally verified for both spectrometers, see \refF{fig:Callibration}. 
The effect is stronger for the LAMPF than for the HPF spectrometer. HPF uses a magnet with an iron yoke confining the stray fields and therefore minimizes the time the muons spend in the external field before implantation in the sample. This yields less spin precession external to the sample compared to LAMPF which uses an air coil.  

When the sample is in the Meissner state it is difficult to fit the parameters $\alpha$ and $f_0$ simultaneously. The strong damping implies that the polarization function never relaxes close to its initial value and $f_0$ and $\alpha$ become strongly correlated. If the sample is in the normal state, $f_0$ equals zero and the polarization function oscillates around zero, see \refF{fig:States}(d). Therefore $\alpha$ can be precisely measured 
in the normal state above $T\msub{c}$ and then be fixed for data obtained below $T\msub{c}$ instead of being used as an additional fit parameter. 

Intuitively, $\alpha$ should not depend on the external field, since it accounts for the detector efficiencies and alignment which should not be affected by the external field. However, experimentally it was found that $\alpha$ changes linearly with field for both spectrometers, see \refF{fig:Callibration}. In the HPF spectrometer the external magnetic field not only acts on the muon spin but also steers the beam before it enters the sample, which can result in a shift of the beam spot and therefore a field dependent $\alpha$. For the LAMPF spectrometer, the field is applied in the direction of muon propagation, see \refF{fig:FieldConfiguration}. However, stray fields, misalignment, and imperfectly polarized beams can still yield a field dependent $\alpha$. As for HPF, a linear, but weaker $\alpha(H)$ dependence was found, see \refF{fig:Callibration}. The linear $\alpha(H)$ relation allows for taking only a few measurements above $T\musb{c}$ and using a linear correction function for calibration. 

In the experiment, the magnetic field is controlled by setting the current $I$ to the magnets. The most accurate way to derive the $B(I)$-relation is using the measurement in the normal state. 

In summary, the normal state calibration serves three purposes:
\begin{enumerate}
	\item{Establish the $B(I)$ relation,}
	\item{Correct for muon precession in the field outside of the sample,}
	\item{Correct for drifts of the beam spot due to steering.}
\end{enumerate}

\begin{figure}[htbp]
    \centering
    \includegraphics[width=\columnwidth]{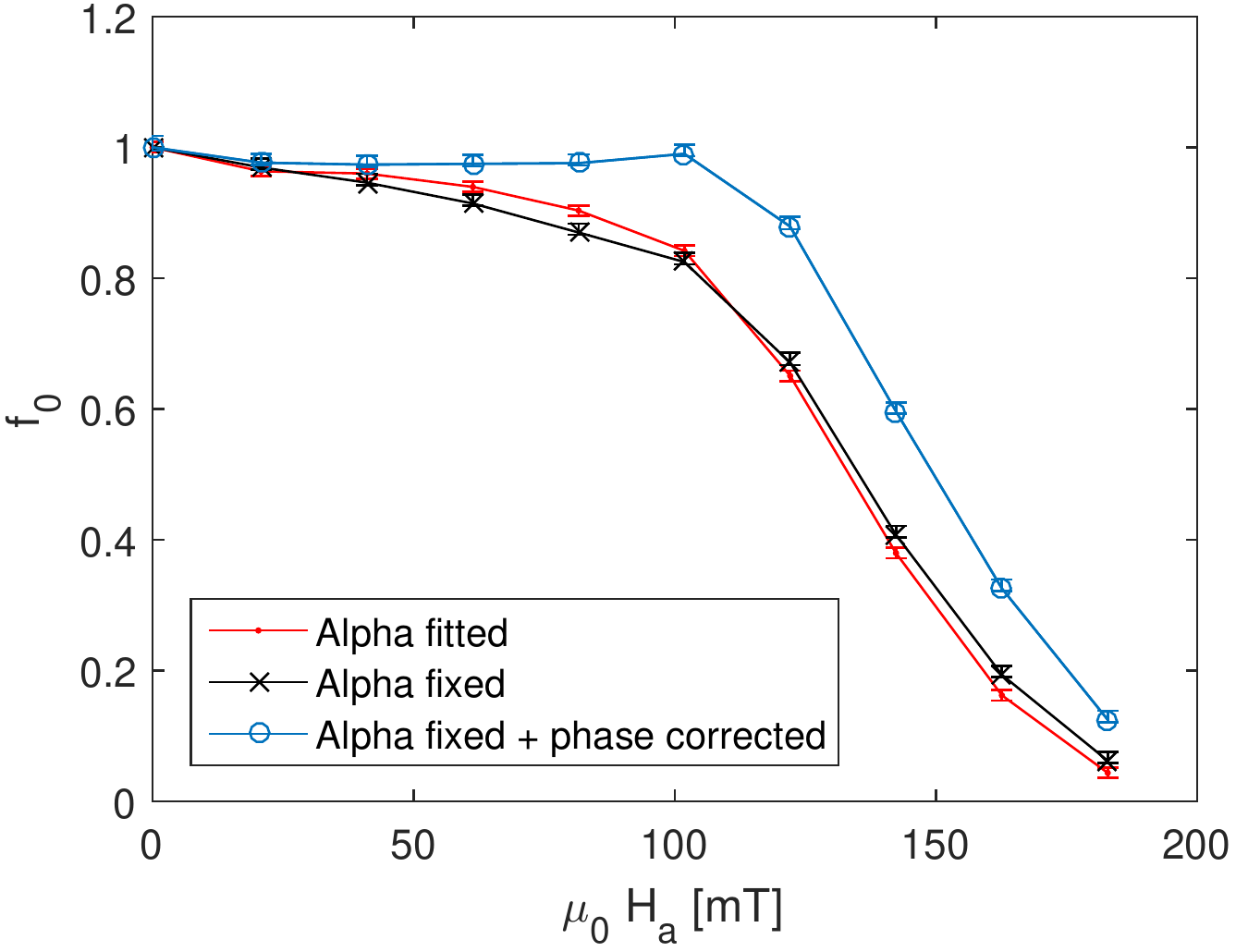}
    \caption{Fit parameter $f_0$ as a function of applied field for a coin sample in transverse geometry (LAMPF). Fixing $\alpha$ from the normal state calibration yields a smother curve, while the phase correction eliminates the effect of reduced initial polarization due to spin rotation outside of the sample.} 
    \label{fig:TR5_callibration}
\end{figure}

Experimentally it was found that a normal state calibration needs to be performed for every sample. While $\Phi (B)$ and $B(I)$ only slightly change for each setup, the critical relation is $\alpha(B)$. Figure\,\ref{fig:TR5_callibration} shows an example of a sample measured on LAMPF. Fixing $\alpha$ yields a smoother curve, especially visible here in the low field area. The phase correction shifts the whole curve up and enables a better estimation of $H\msub{a$\mid$entry}$, effectively eliminating the effect of muon spin rotation outside of the sample.

%


\subsection{Samples}
\begin{figure}[tbh]
   \centering
	 \includegraphics[width=0.9\columnwidth]{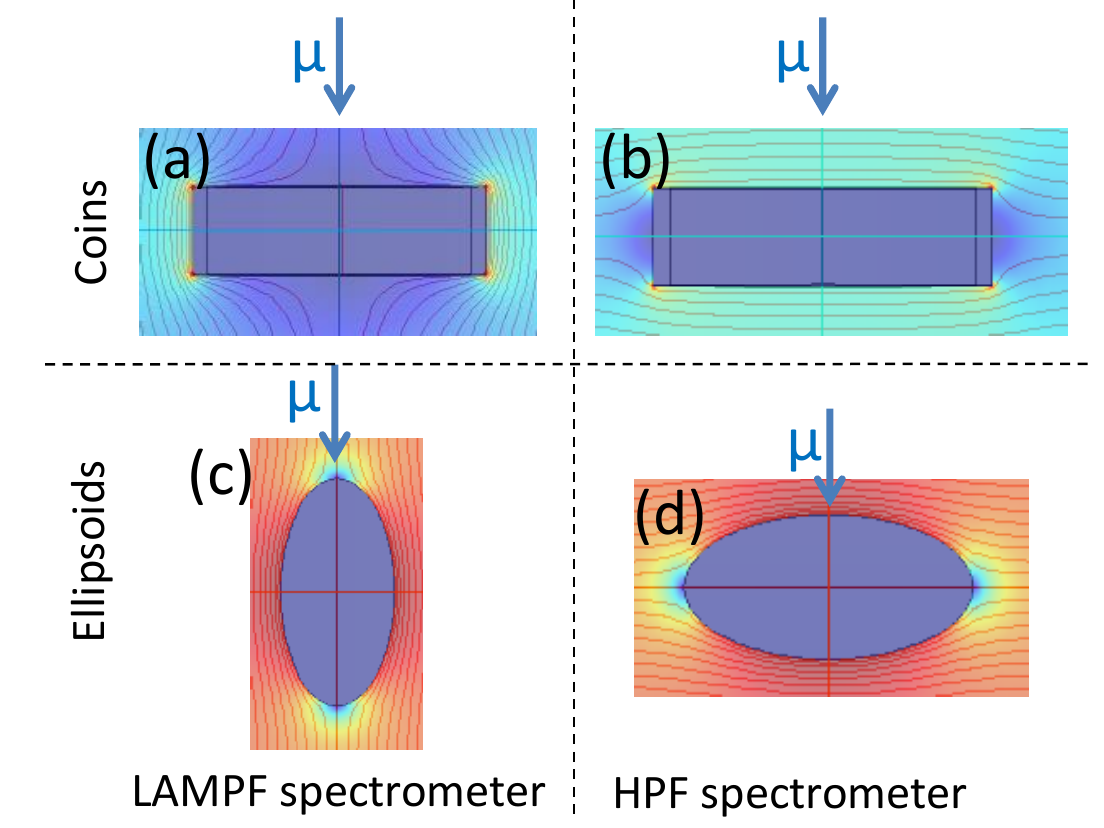}
   \caption{Four generic arrangements of sample, muon and field direction using the LAMPF and HPF spectrometers. The direction of the muon spin is always perpendicular to the beam direction and the sample surface. Note the high flux line density at the edges of the coin in the transverse geometry (LAMPF). In the parallel geometry (HPF) the field enhancement at the edges is much weaker. For the ellipsoid there is no edge boundary. Here the field will first nucleate at the equator, where the muons are implanted in the HPF setup.}
	\label{fig:FieldConfiguration}
\end{figure}

Several sample types and fields geometries are used. Unless otherwise stated, all samples are made of RRR niobium, which specifies niobium with a RRR$>$300. Coin samples of \unit[3]{mm} thickness and \unit[20]{mm} diameter are cut by water jet from flat sheets. Similar coins were cut by wire Electrical Discharge Machining (EDM) from a \unit[1.3]{GHz} cavity half-cell of TESLA shape at a location \unit[45]{\degree} from the equator as rotated toward the iris. This half-cell was made by deep drawing from a sheet of \unit[3]{mm} thickness. These cylindrical samples can be tested in parallel and perpendicular field geometry. Figure \ref{fig:FieldConfiguration}(a) displays the initial perpendicular field configuration, while Fig. \ref{fig:FieldConfiguration}(b) shows the direction of applied magnetic field and muon propagation for the HPF spectrometer developed to test samples in a parallel field geometry. 

Another set of samples were machined in the shape of a prolate ellipsoid. The dimensions are semi-major axis of \unit[22.9]{mm} and semi-minor circular cross-section of \unit[9.0]{mm} radius. Moreover, along the major axis, at one end there is a \unit[21]{mm} deep 1/4-20 threaded hole which was used to hold the sample. These samples can be tested in the initial LAMPF spectrometer. Here, the magnetic field is applied along the major axis of the sample and the muons are being implanted on the tip of the sample, see \refF{fig:FieldConfiguration}(c). A fourth set of samples consists of smaller ellipsoids which can be used in the HPF spectrometer, see \refF{fig:FieldConfiguration}(d).  These ellipsoids have the same aspect ratio as the larger ones but have been scaled down to a semi major axis length of \unit[16]{mm} to fit the cryostat. To hold the sample, a 4/40 threaded hole is placed on the minor axis. 

\begin{figure}[tbh]
   \centering
	 \includegraphics[width=0.9\columnwidth]{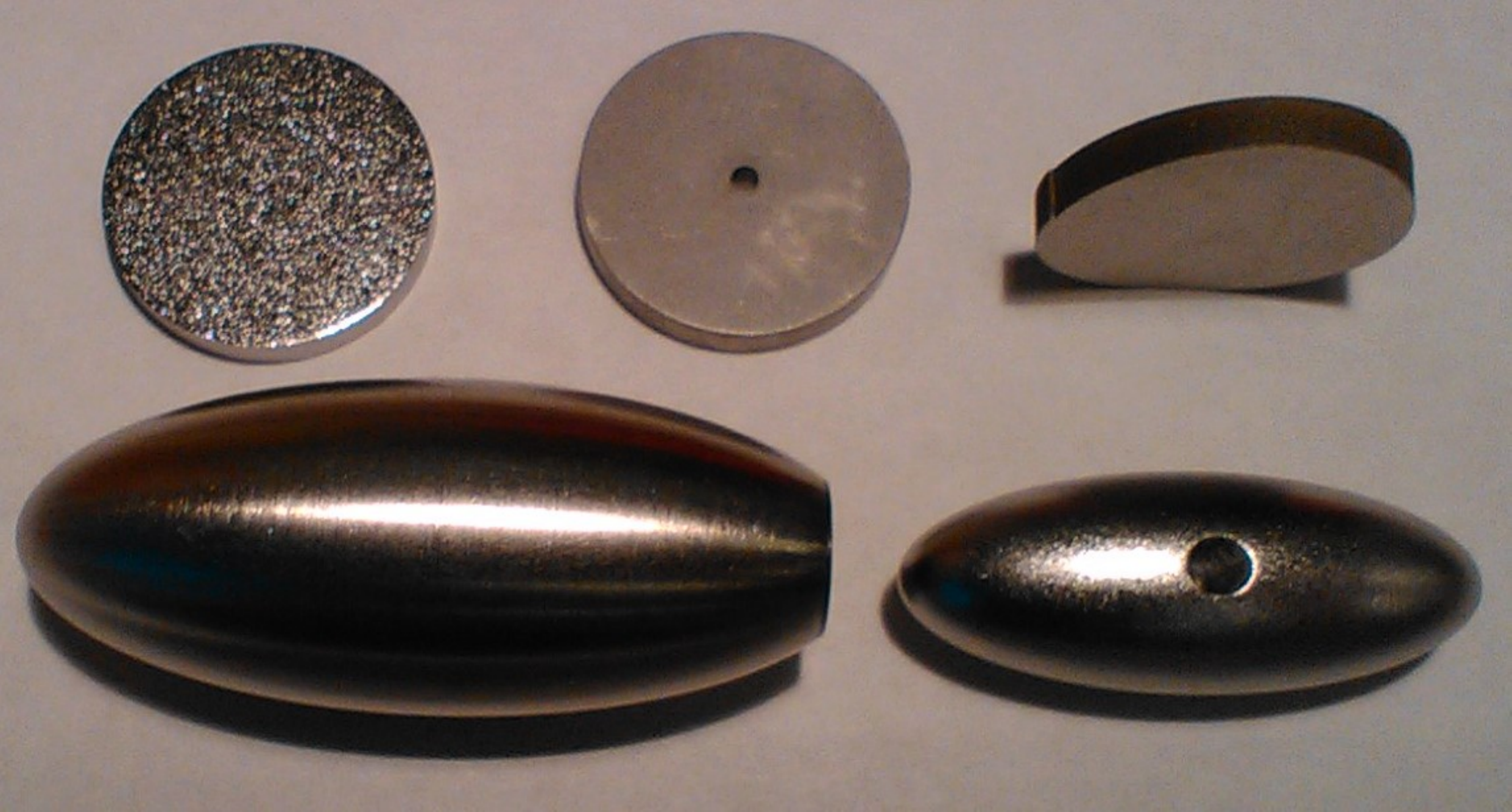}
   \caption{Example of samples used in the experiment. Top, from left to right: Coins after BCP, after deposition of Nb3Sn (the whole in the middle was used to hang the sample in the furnace) and cut from a \unit[1.3]{GHz} cavity half cell. Bottom: Ellipsoids used in LAMPF (left) and HPF (right).}
	\label{fig:Samples}
\end{figure}

The samples were subjected to a variety of different treatments typical for SRF cavity processing. These included heat treatments in vacuum such as \unit[120]{{\degree}C} bake for \unit[48]{hours} and \unit[800]{{\degree}C} degassing for \unit[4]{hours}. Vacuum heat treatments at \unit[1200]{{\degree}C} and \unit[1400]{{\degree}C} for \unit[4]{hours} each were also employed. Surface treatments include etching using both buffered chemical polish (BCP) and electro-polish (EP) with various removals. To study materials other than niobium some samples were coated. For example, Fig. \ref{fig:Samples} displays a coin which has received a Nb3Sn coating using a thermal diffusion technique at Cornell University.  

\section{Effect of pinning and geometry}
\subsection{Coins in transverse geometry}

Consider first the coin sample with field applied perpendicular to the face (\refF{fig:FieldConfiguration}(a)). When in the Meissner state, surface currents will be set up to cancel the field in the bulk. The magnetic field will be enhanced at the edges of the coin by a factor related to a demagnetization factor $N$ by $H\msub{edge}=H\msub{a}/(1-N)$ where $H\msub{a}$ is the applied field. In the literature, $N$ is often more specifically referred to as the magnetometric demagnetization factor to distinguish it from the fluxmetric (also known as ballistic) demagnetization factor $N\msub{s}$. The latter is related to flux penetration into the midplane of the samples. For non-elliptical shapes $N$ and $N\msub{s}$ both depend not only on the sample geometry but also on the susceptibility of the material $\chi$ \cite{chen2006fluxmetric}. Numerical calculations of $N$ and $N\msub{s}$ necessarily require a constant $\chi$. For superconductors, $\chi=-1$ is only valid in case of complete shielding. Therefore, this concept is applicable to calculate $H\msub{edge}$ from $N$, but not $H\msub{entry}$ from $N\msub{s}$. For the latter, Brandt has developed a model which calculates magnetization curves $M(H\msub{a})$ for some geometries and derives $H\msub{entry}$ from its maximum.

For Type II superconductors when the applied field is such that the enhanced field at the edges reaches $H\msub{c1}$, the field will break into the edge such that the local field is reduced due to the rounding of the flux line. As the field increases the flux lines will cut further across the corner and eventually join at the center of the sample edge. This corresponds to $H_{\text{a}\mid \text{entry}}=H\msub{a}/(1-N\msub{s})$ and is higher than $H\msub{c1}\cdot(1-N)$ due to the so called edge boundary \cite{brandt2000superconductors}. The flux line now crosses the full sample width and is driven inwards due to interaction with the surface currents.
\begin{figure}[tbh]
   \centering
 \includegraphics[width=0.7\columnwidth]{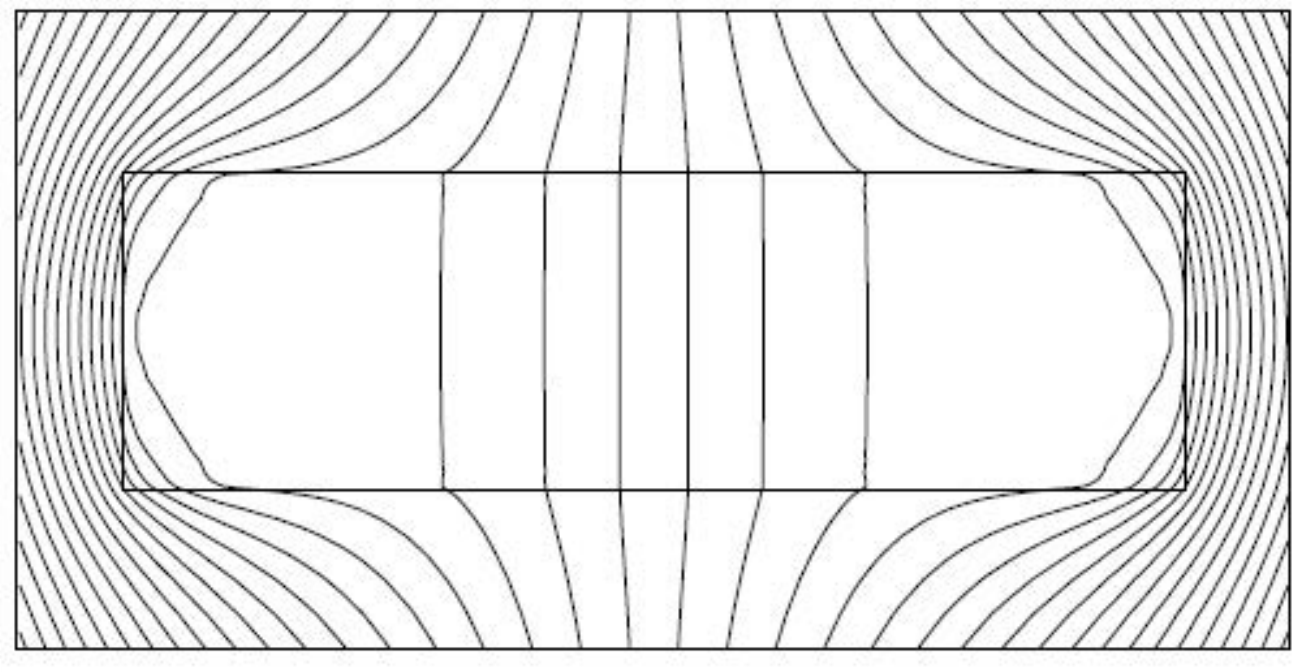}
   \caption{Flux applied to a thin circular disk transverse to an applied field where $H\msub{a}>H\msub{a$\mid$ entry}$. The field breaks in at the edges first at $H\msub{edge}<H\msub{a$\mid$ entry}$. Above $H\msub{a$\mid$ entry}$ the flux lines will move to the center of the sample, which is the position with lowest line tension. Reproduced from \cite{brandt2000superconductors}. }
	\label{fig:brandtflux}
\end{figure}

In a pin-free sample the flux will move to the center since this represents the lowest energy position (minimum line tension), see \refF{fig:brandtflux}. As the flux increases and vortices multiply, the vortex currents will repel so that the flux lines will redistribute and fill from the center to the outside edge. In our case for the transverse coin geometry (cylinder in axial field), with diameter $a$=\unit[20]{mm} and thickness $b$=\unit[3]{mm}, the demagnetizing factor is $N$=0.77 meaning that $H\msub{a$\mid$ edge}=0.23H\msub{c1}$. 
Brandt \cite{brandt2000superconductors} derives the field of first flux entry (to the midplane) from the maximum $M(H\msub{a})$, where $M$ is the magnetization. For a cylinder in an axial field he finds:  

\begin{equation}
H\msub{a$\mid$ entry}=\tanh{\sqrt{0.67\frac{b}{a}}\cdot H\msub{c1}=0.31H\msub{c1}}.
\label{eq:Hentry}
\end{equation}

\begin{figure}[htbp]
    \includegraphics[width=\columnwidth]{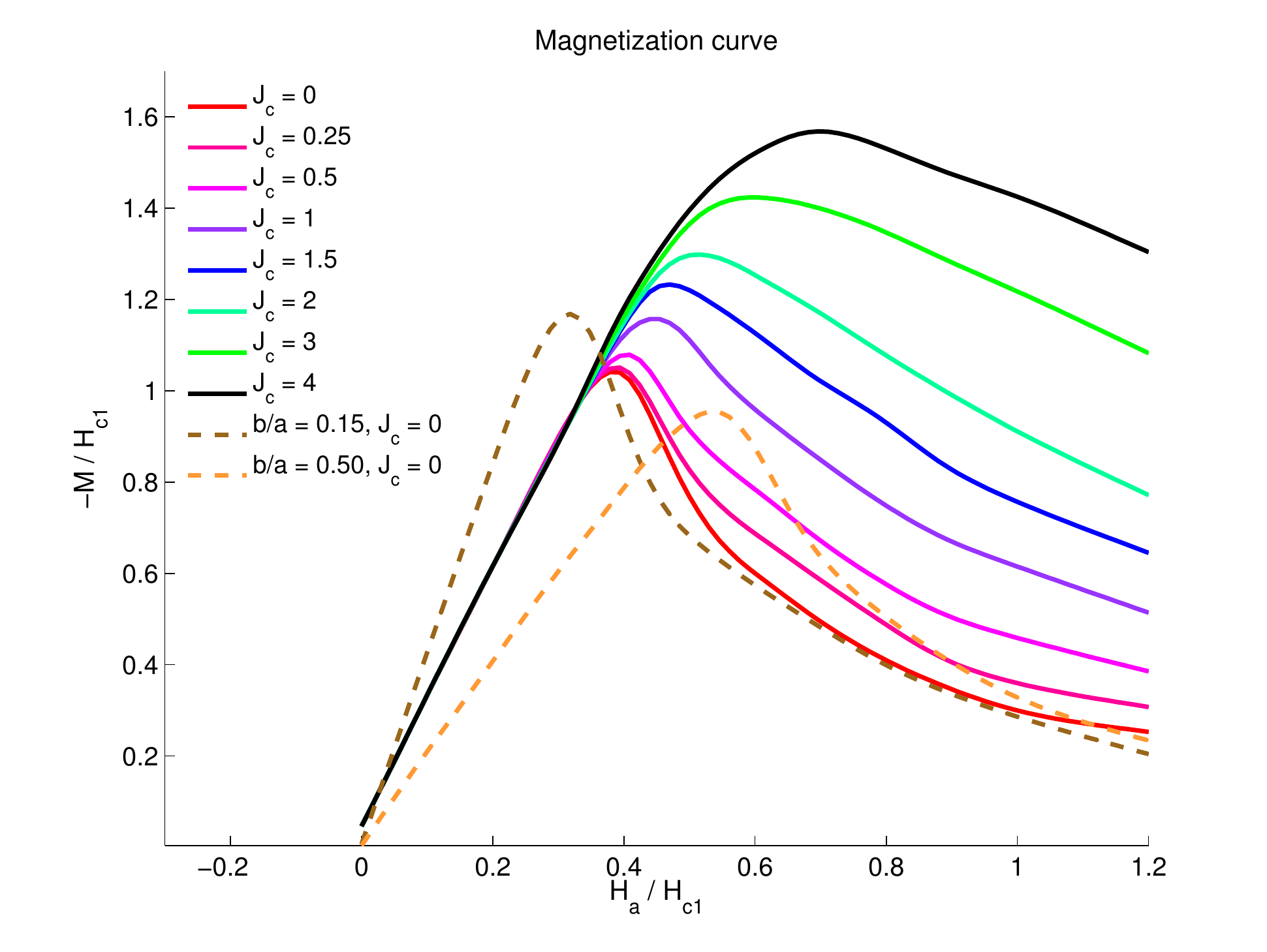}
    \caption{Magnetization curves for a disk sample with different pinning and geometry. Solid lines correspond to the sample with b/a = 0.3 while the dash line correspond to pin-free sample with b/a = 0.15 and 0.50. The critical current density $J\msub{c}$ is a measure of the pinning strength. The larger $J\msub{c}$ the stronger the pinning. Adapted from \cite{brandt2000superconductors}. }
    \label{fig:PinnedMagCurve}
\end{figure}

For a sample with pinning, the pinning centers act as additional barriers adding ‘resistance’ to the mobility of vortices moving from the edges to the center and increasing $H\msub{a$\mid$ entry}$ compared to the pin free case, see \refF{fig:PinnedMagCurve}. Hence, introducing pinning into the material delays the entry of magnetic field into the center of the sample. 

\subsection{Parallel coin}
\label{sec:Parallel coin}
In the parallel geometry (\refF{fig:FieldConfiguration} b), the sample coin is placed parallel to the applied field and the muons are applied to the coin face. The demagnetization factor in this radial geometry has been calculated by Chen et al. \cite{chen2001radial}. For our standard geometry (diameter $a$=\unit[20]{mm} and thickness $b$=\unit[3]{mm}), $N$=0.15. To estimate $H\msub{a$\mid$entry}$ we cannot use a literature value of a fluxmetric demagnetization factor, since this concept relying on a constant $\chi$ is not applicable as mentioned above. Furthermore, for this radial geometry an approximation formula for $H\msub{a$\mid$entry}$ has not yet been derived to our knowledge. Brandt in \cite{brandt2000superconductors} derives a formula for a long strip with rectangular cross section $a\times b$ with the field applied along $a$,
\begin{equation}
H\msub{a$\mid$ entry}=\tanh{\sqrt{0.36\frac{a}{b}}\cdot H\msub{c1}}.
\label{eq:Hentry2}
\end{equation}
Since we are only interested in the inner area probed by the muons (\unit[8]{mm} diameter) this geometry should be applicable to our setup. This assumption will be reviewed later by comparing samples of different geometry and identical preparation. For our standard geometry (diameter $a$=\unit[20]{mm} and thickness $b$=\unit[3]{mm}) we find $H\msub{a$\mid$entry}=0.91H\msub{c1}$. 



In this parallel geometry, the volume sampled by the muons is less sensitive to pinning. Flux could still be pinned at the corners before linking at the center (pinning enhanced edge boundary) but much less so than in the transverse geometry since no flux motion from the edges of the sample to its center is required as in the transverse geometry.

\subsection{Ellipsoids}
For the ellipsoidal geometry the edge boundary is eliminated. The inward directed driving force on the vortex ends by the surface screening currents is compensated by the vortex line length that increases for fluxoids that are closer to the ellipsoid axis – so pin-free ellipsoidal samples produce a uniform vortex flux density in the mixed state. The Meissner state is supported by screening currents that augment the field at the equator and reduce the field at the poles. When the flux at the equator reaches $H\msub{entry}$, which could be either the lower critical field $H\msub{c1}$ or in case of a surface barrier the superheating field $H\msub{sh}$, fluxoids will nucleate at the equator and redistribute uniformly inside the superconductor due to vortex repulsion for a pin free sample. 


In our geometry, the demagnetizing factor is $N$=0.13 with $H\msub{a$\mid$entry}=0.87H\msub{entry}$ \cite{brandt2000superconductors}, where $H\msub{entry}$ denotes the intrinsic field of first entry of the material. In the case of samples with pinning, the redistribution will be affected as the pinning centers will add a frictional component to the redistribution such that the fluxoids will tend to preferentially populate nearer the equator and will only gradually reach the poles as the applied field increases beyond $H\msub{a$\mid$entry}$. 

The parallel field ellipsoid geometry with an application of muons at the equator (Fig. \ref{fig:FieldConfiguration}(d)) should be the least sensitive to pinning and is our preferred geometry to probe the intrinsic field of first flux entry $H\msub{entry}$.

\subsection{Coated samples}
Consider now a niobium sample coated with a thin layer of a material with larger $T\msub{c}$. If this layer is thinner than the implantation depth of the muons and measured above $T\msub{c}$ of niobium but below $T\msub{c}$ of the coating, the geometry will be a superconducting shell.

With the bulk of the sample being normal conducting and therefore not providing any pinning, the geometric boundary is eliminated since as soon as flux breaks into the corners the fluxoid will snap to the center for a pin free shell. For the case of the transverse coin $H\msub{a$\mid$entry}=H\msub{a$\mid$edge}=0.23H\msub{c1}$, while for the superconducting shell in parallel geometry we expect $H\msub{a$\mid$entry}=0.85 H\msub{c1}$. For ellipsoidal shells the situation is similar to bulk ellipsoids in terms of magnetization except that after nucleation, the fluxoids will snap to the center since the flux line length in the superconducting shell is actually less near the ellipsoid axis so that pinning would be less dominant in resisting nucleated flux to move to the poles. Table\,\ref{tab:geometry} displays the demagnetization factor $N$ and $H\msub{a$\mid$entry}/H\msub{c1}$ for all geometries. Note that in case of the ellipsoids, the field direction with respect to the sample geometry is identical in the two spectrometers and only the muon implantation site is changed. Therefore, $N$ and $H\msub{a$\mid$entry}/H\msub{c1}$ are identical for the two ellipsoid arrangements. 

\subsection{Expected field of first entry}
Measurements are typically performed at about \unit[2.5]{K}. For the critical temperature of niobium \unit[9.25]{K} \cite{finnemore1966superconducting} and assuming the empirical relation for the temperature dependance 
\begin{equation}
H\msub{c1}(T)=H\msub{c1}(0)\left(1-\left(\frac{T}{T\msub{c}}\right)\right)
\label{eq:Hc1(T)}
\end{equation}  
$\mu_0 H\msub{c1}$(2.5K)$\approx$\unit[161]{mT} can be obtained, assuming $\mu_0 H\msub{c1}(0K)=\unit[174]{mT}$ \cite{finnemore1966superconducting}. Finally, the expected field of first entry for a pin free niobium sample with no surface barrier 
\begin{equation}
H_0(T)=\frac{H\msub{a$\mid$entry}}{H\msub{c1}} H\msub{c1}(T) 
\end{equation}
can be calculated. Table \ref{tab:geometry} displays $\mu_0 H_0(2.5K)$ for all geometries.



\begin{table}
	\centering
		\begin{tabular}{  l | c | c | c | c } 
Sample	& SC type	& $N$ &	$H\msub{a$\mid$entry}/H\msub{c1}$ &$\mu_0 H_0$(2.5K)[mT] \\ \hline
Transverse coin	& bulk	& 0.77	& 0.31 & 50 \\
Parallel Coin	& bulk	& 0.15	& 0.91 &146\\
Ellipsoid	& bulk	& 0.13	& 0.87 &140 \\
Transverse coin	& shell	& 0.77	& 0.23 &37\\
Parallel Coin	& shell	& 0.15	& 0.85 & 137\\
Ellipsoid	& shell	& 0.13	& 0.87 & 140\\
\end{tabular}
\caption{Geometrical normalizing factors for expected pin-free flux entry for the three sample types assuming $\mu_0 H\msub{c1}(0K)$=\unit[174]{mT}. For the shell geometry the edge boundary is eliminated yielding $H\msub{a$\mid$entry}=H\msub{edge}=(1-N)H\msub{c1}$.}
\label{tab:geometry}
\end{table}

\section{Results}

\subsection{Comparing Geometries}
\label{Comparing Geometries}

We present several results from different samples first to illustrate the effect of the geometry. In these and subsequent plots the field is normalized to $H/H_0$ where $H_0$ corresponds to the expected entry field for a pin-free sample, with demagnetization and edge boundary considered, assuming $H\msub{c1}(0K)$=\unit[174]{mT} \cite{finnemore1966superconducting}. 
For the critical temperature of niobium \unit[9.25]{K} and assuming the empirical relation for the temperature dependance \refE{eq:Hc1(T)},
$H\msub{c1}$(T) is obtained individually for each curve. Table\,\ref{tab:geometry} gives the estimated $H_0$ values for the three geometries at the typical measurement temperature \unit[2.5]{K}. 
Statistical fit errors are on the order of the size of the markers or smaller, see \refF{fig:TR5_callibration}, and are for simplicity not displayed in subsequent plots. The resolution of the $H\msub{entry}$ measurement is determined by the number of data points $f_0(H)$ taken in the area of transition from the Meissner to the vortex state. The uncertainty for the measured $H$ is mostly affected by misalignment, which is estimated to be below \unit[5]{$\degree$} corresponding to less than \unit[1]{mT} even at high field. 
%

\begin{figure}[htbp]
    \centering
    \includegraphics[width=\columnwidth]{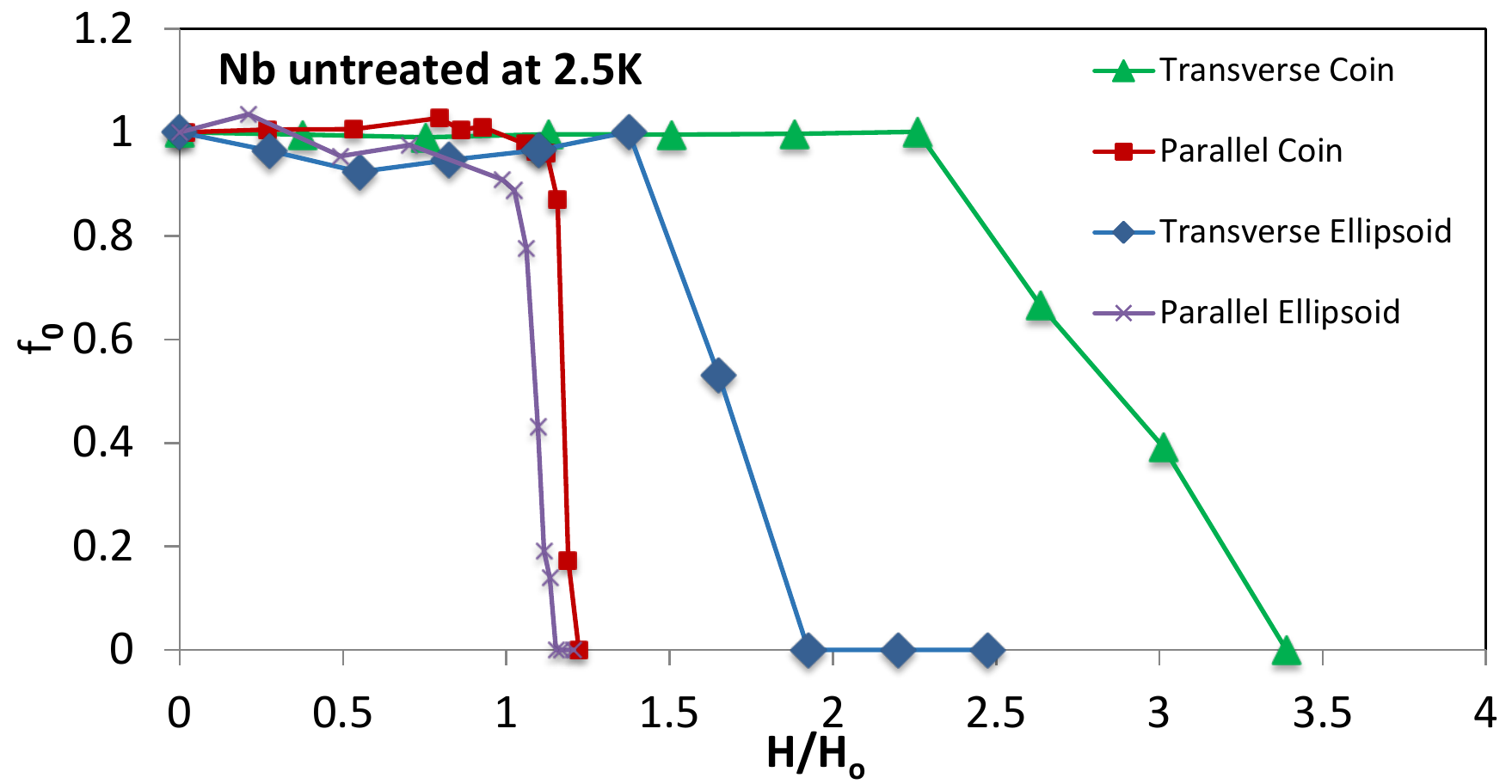}
    \caption{Fit parameter $f_0$ signifying the volume fraction probed by the muons which is in the field free Meissner state as a function of applied field for chemically etched samples with no heat treatment in four geometries. The apparent differences in $H/H_0$ are correlated to the difference sensitivity to pinning of the four geometries.}
    \label{fig:UntreatedSamples}
\end{figure}

Figure \ref{fig:UntreatedSamples} shows $f_0$ as a function of applied field for the four different sample/field arrangements used in this experiment. All these samples have received buffered chemical polishing (BCP), but no bakeout. The pinning strength is therefore expected to be rather strong. For all samples, $f_0$ stays above 0 beyond the expected field of first entry. The effect is most strongly pronounced for the coin in transverse geometry. In comparison, ellipsoid shaped samples in the same spectrometer (LAMPF) yield a geometry less sensitive to pinning. The HPF spectrometer is better suited for field of first entry measurements. Here, $f_0$ reaches a 0 value at a field closer to the predicted value for niobium of $H/H_0=1$. 

\begin{figure}[htbp]
    \centering
    \includegraphics[width=\columnwidth]{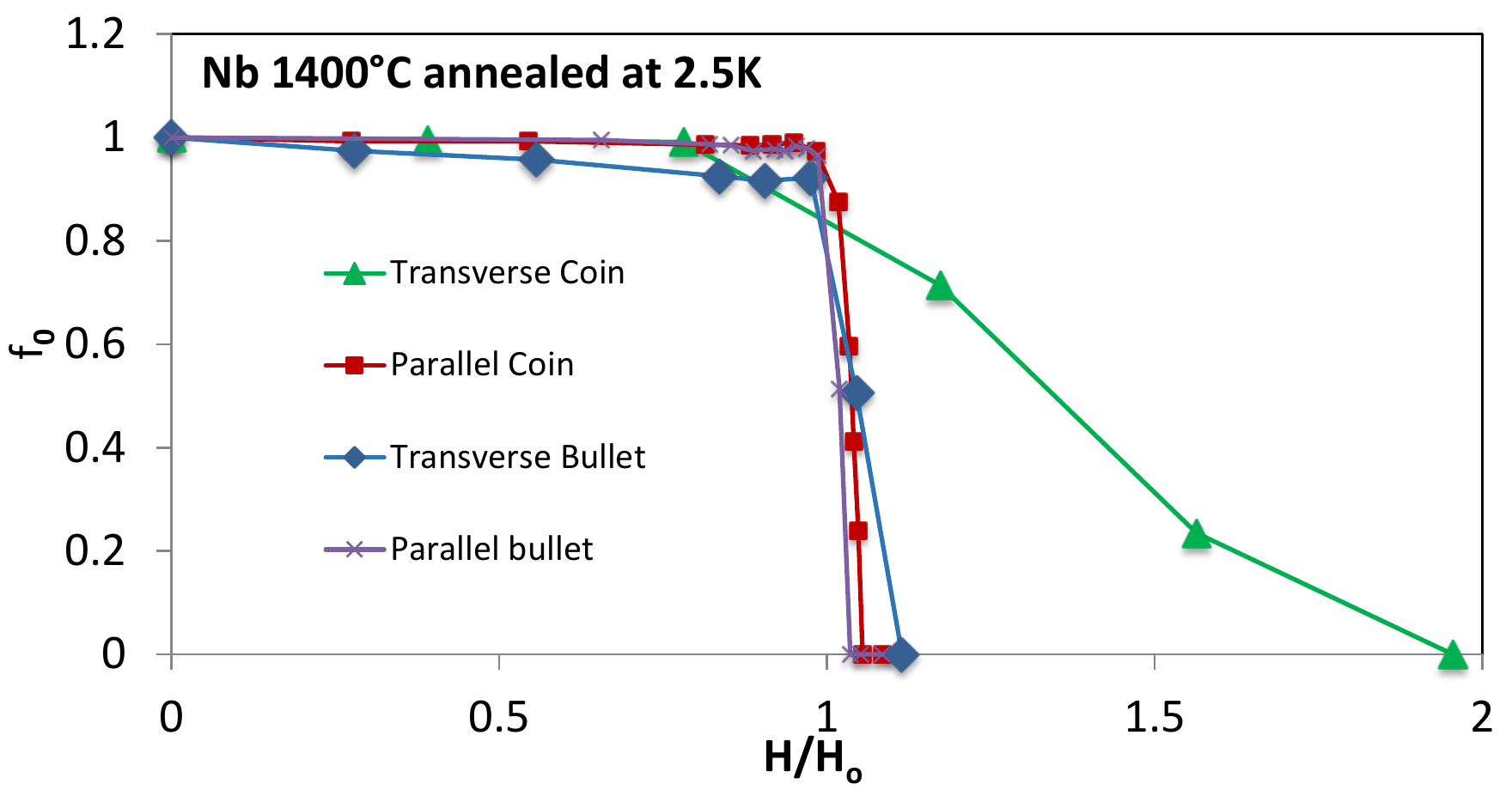}
    \caption{Fit parameter $f_0$ signifying the volume fraction probed by the muons which is in the field free Meissner state as a function of applied field for samples annealed at \unit[1400]{$\degree$C} in four geometries. The annealing virtually releases all pinning.}
    \label{fig:1400C}
\end{figure}

In \refF{fig:1400C} the samples are heat treated at \unit[1400]{$\degree$C}. This virtually eliminates all pinning as the material is fully recrystallized at this temperature. In principal, it is possible to estimate $H\msub{entry}$ even in the transverse coin geometry for these fully annealed samples. However, it is evident from the plot that the interpretation of the data is less ambiguous in all other geometries due to the geometrical edge boundary that delays flux entry to the center. For these pin-free samples we find $\mu_0H\msub{entry}$=\unit[176(4)]{mT} for the parallel bullet and \unit[179(3)]{mT} for the parallel coin. These values are identical within the resolution of the measurement and therefore confirm that the geometrical approximation (Sec.\,\ref{sec:Parallel coin}) of the parallel coin as a long strip is actually applicable here. Both values are close to $H\msub{c1}(0K)$=\unit[174]{mT} as reported by Finnmore \cite{finnemore1966superconducting}. 

In order to investigate materials with unknown $H\msub{entry}$ and pinning strength, the coin shape is ideal. It requires however that the sample is tested with both spectrometers. First the sample can be measured in parallel geometry and the field where $f_0$ deviates from 1 is interpreted as $H\msub{entry}$. The sample needs to be subsequently measured in transverse geometry to estimate whether the obtained $H\msub{entry}$ has been overestimated due to strong pinning.


\subsection{Transverse Coin Results}
The results presented so far show that the transverse geometry is especially sensitive to pinning in the sample. In the following, the test results will be used to show how various surface and bulk treatments can affect the pinning strength. For the study of surface treatment and pinning strength, five samples were cut out from the same RRR niobium sheet. One sample received no further treatment while the other were chemically etched (BCP) to remove \unit[100]{$\mu$m} material. Three of these samples were subsequently baked at \unit[120]{$\degree$C} for \unit[48]{hours}. Afterwards two samples received additional surface treatments, one a \unit[5]{$\mu$m} BCP and the other one a rinsing with hydrofluoric acid to remove and regrow the oxide layer. The results show that the pinning strength is reduced by the initial BCP as the original damaged layer is removed. Further, it can be observed that low temperature baking and surface treatments after BCP have no effect, indicating that beyond the gross removal of surface pollution, the pinning is more dependent on the bulk properties. 

\begin{figure}[htbp]
    \centering
    \includegraphics[width=\columnwidth]{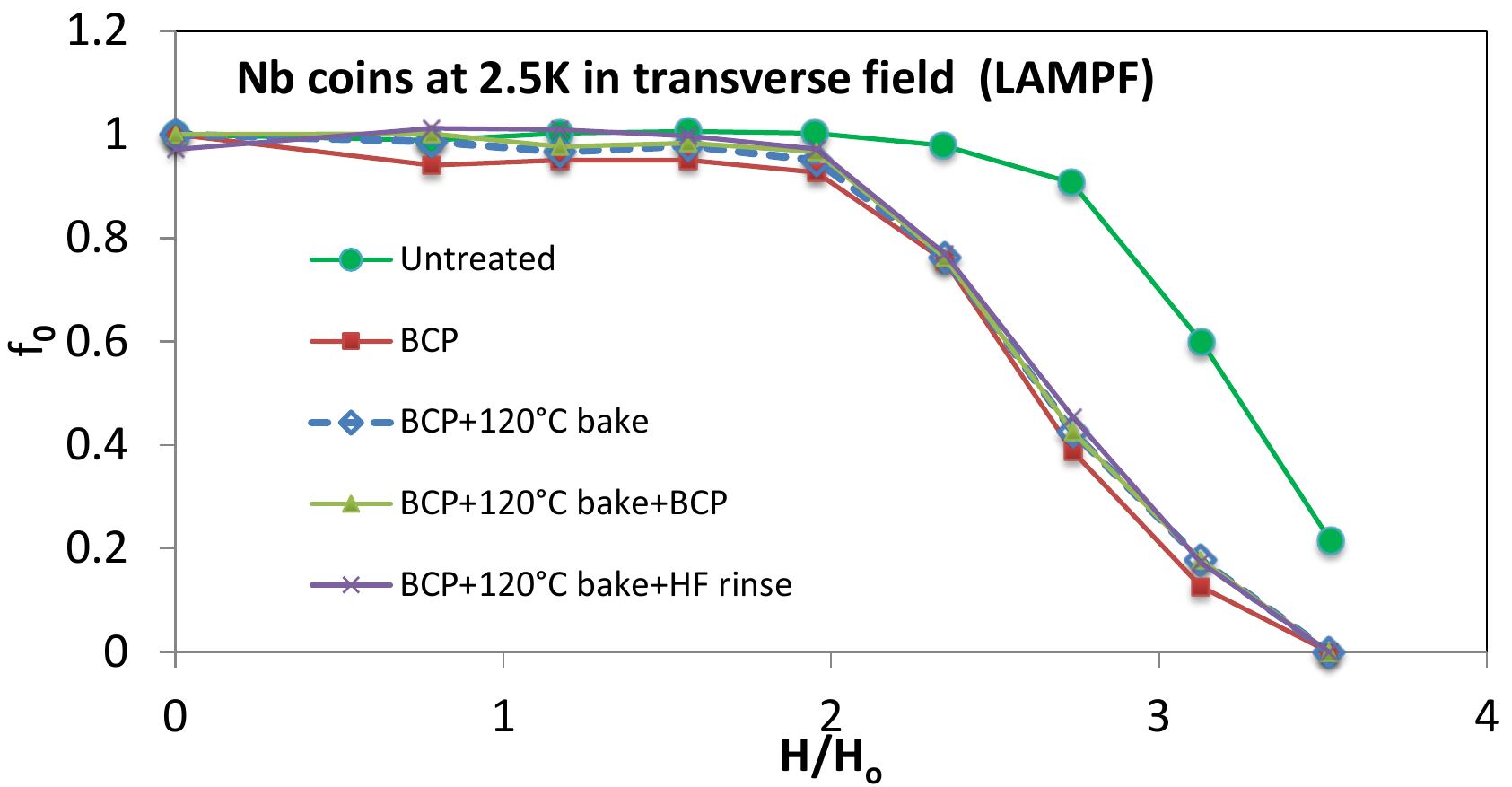}
    \caption{Fit parameter $f_0$ signifying the volume fraction probed by the muons which is in the field free Meissner state as a function of applied field for coins in transverse geometry with different surface treatments at \unit[2.5]{K}. BCP removes the outer demaged layer and releases pinning. Additional surface treatments have no effect on the pinning strength.}
    \label{fig:120C}
\end{figure}		
  
\begin{figure}[htbp]
    \centering
    \includegraphics[width=\columnwidth]{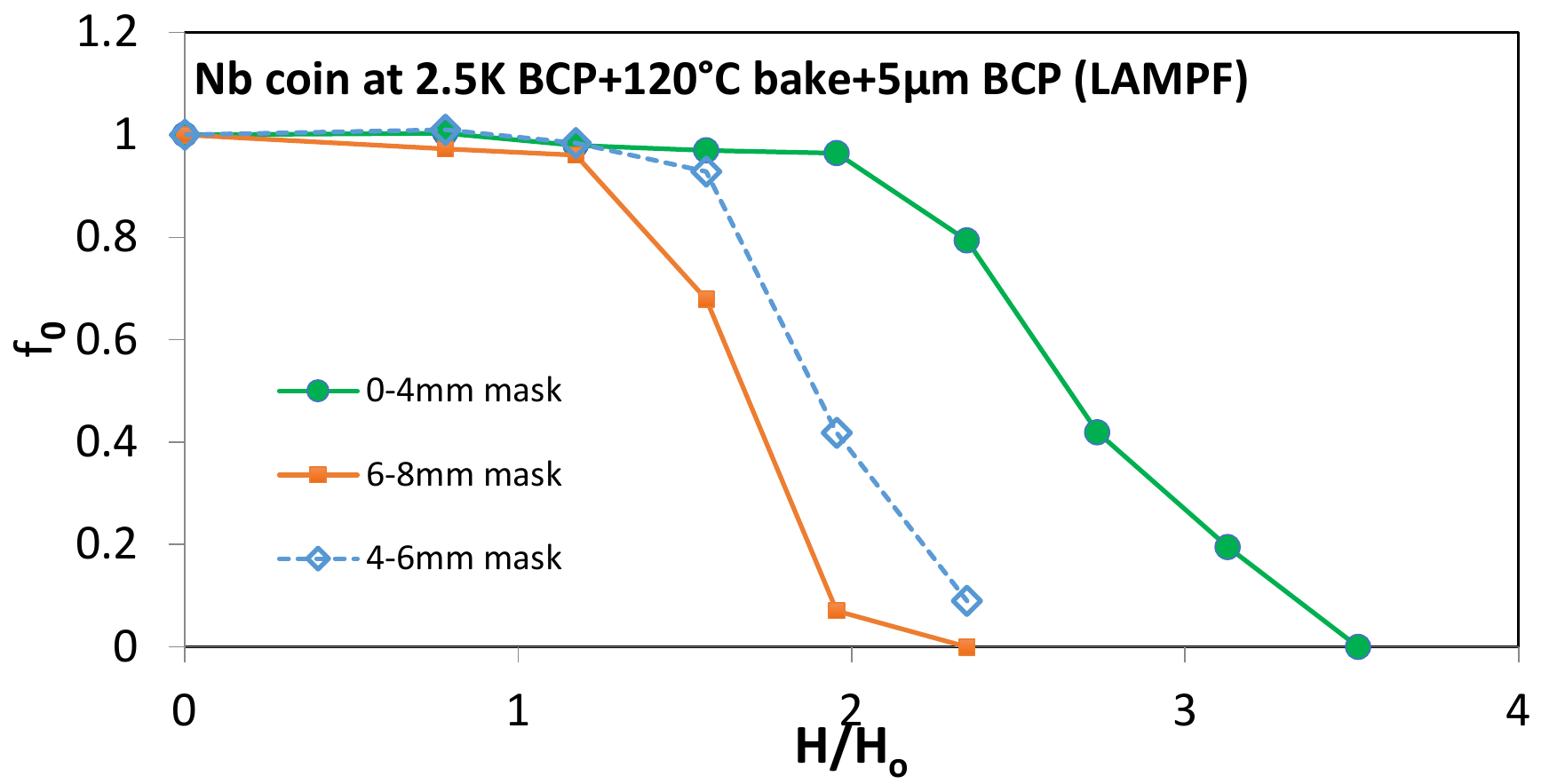}
    \caption{Fit parameter $f_0$ signifying the volume fraction probed by the muons which is in the field free Meissner state as a function of applied field with different masks, probing different areas on the sample visualizing how flux breaks in from the corner of the sample in the transverse coin geometry. }
    \label{fig:annular}
\end{figure}

To get more information on how the flux breaks in for the case with pinning, a series of measurements were taken with different masking foils: a standard \unit[8]{mm} aperture (\unit[4]{mm} radius), an annular mask blocking the center of the sample with an inner and outer radius of 4 and \unit[6]{mm}, and a second annular mask with radii \unit[6-8]{mm}. The results are plotted in \refF{fig:annular}, once again with fields normalized to the expected $H\msub{a$\mid$entry}$ based on the geometry. The sample used for this test was first treated by a bulk buffered chemical polishing (BCP) removing \unit[100]{$\mu$m}, followed by \unit[120]{$\degree$C} baking in vacuum and a final \unit[5]{$\mu$m} BCP. For this treatment the pinning is rather strong as can be seen from the curve obtained with the standard \unit[0-4]{mm} mask, also displayed in \refF{fig:120C}. The flux is not driven to the center until the field reaches over \unit[two]{times} the pin-free $H\msub{a$\mid$entry}$, and is not fully saturated until over\unit[three]{times} $H\msub{a$\mid$entry}$. Comparing results obtained with the different masks, we find that the field breaks in near $H\msub{a$\mid$entry}$ at large radii but does not migrate to the center as would be expected for a pin free case. The results are a strong confirmation of the role of pinning as a source of flux drag that inhibits redistribution of the penetrating flux into the sample center.

\begin{figure}[htbp]
    \centering
    \includegraphics[width=\columnwidth]{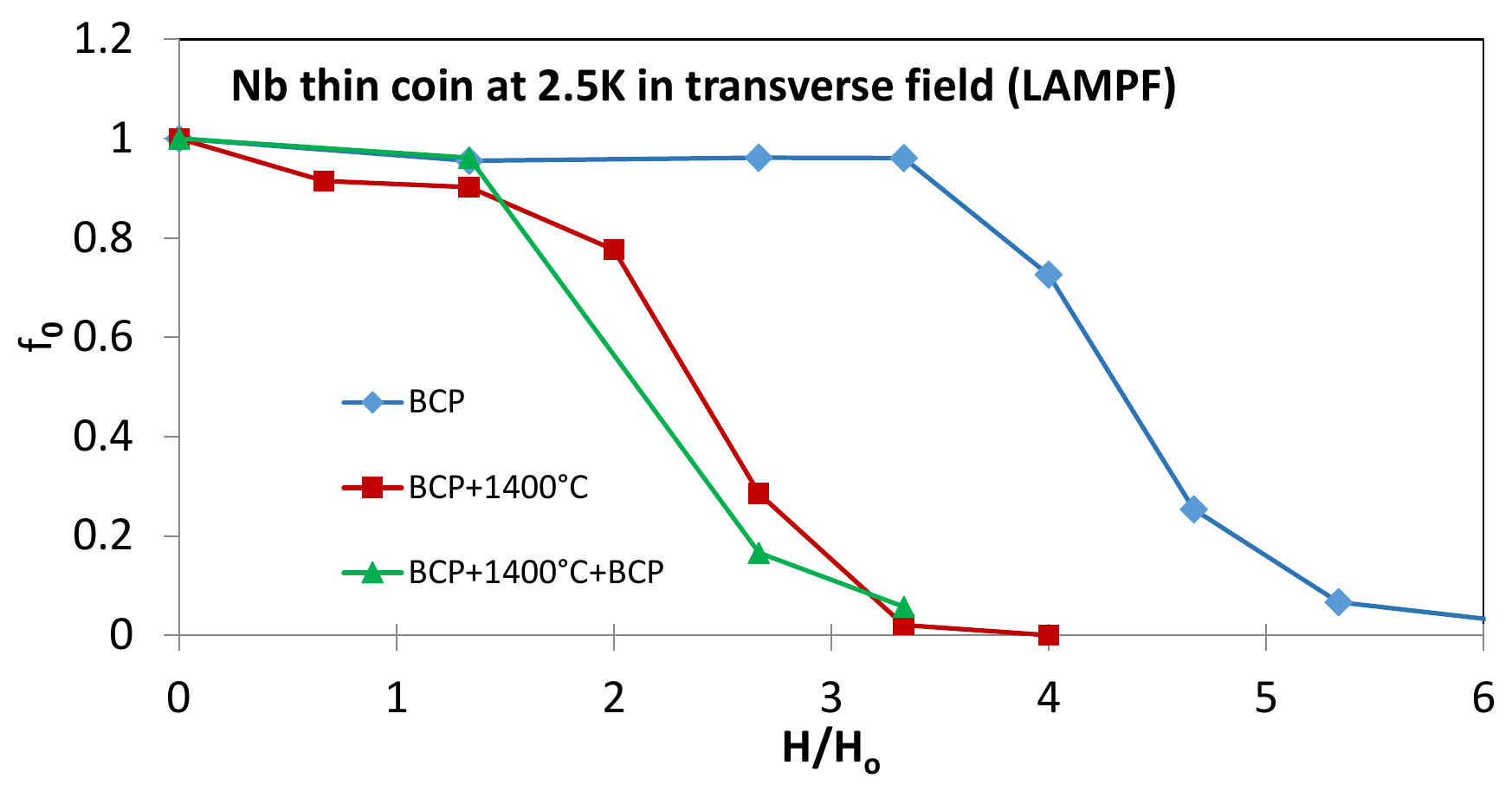}
    \caption{Fit parameter $f_0$ signifying the volume fraction probed by the muons which is in the field free Meissner state as a function of applied field for a \unit[0.8]{mm} thin sample of \unit[18]{mm} diameter. The sample was first etched (BCP), followed by annealing (1400$\degree$C) and another BCP. This plot demonstrates that pinning is essentially a bulk rather than a surface effect, since a BCP treatment after annealing does not increase the pinning strength. }
    \label{fig:thin_coin}
\end{figure}

Figure \ref{fig:thin_coin} shows results from a thinner coin cut from RRR Nb. This sample was first etched (BCP) and demonstrates the characteristics of strong pinning. The sample was then heat treated at \unit[1400]{$\degree$C}, resulting in a significant decrease in pinning. When the annealed sample is etched again, removing \unit[7]{$\mu$m} material, the pinning did not return showing that the pinning is predominantly a bulk rather than a surface effect.

\begin{figure}[htbp]
    \centering
    \includegraphics[width=\columnwidth]{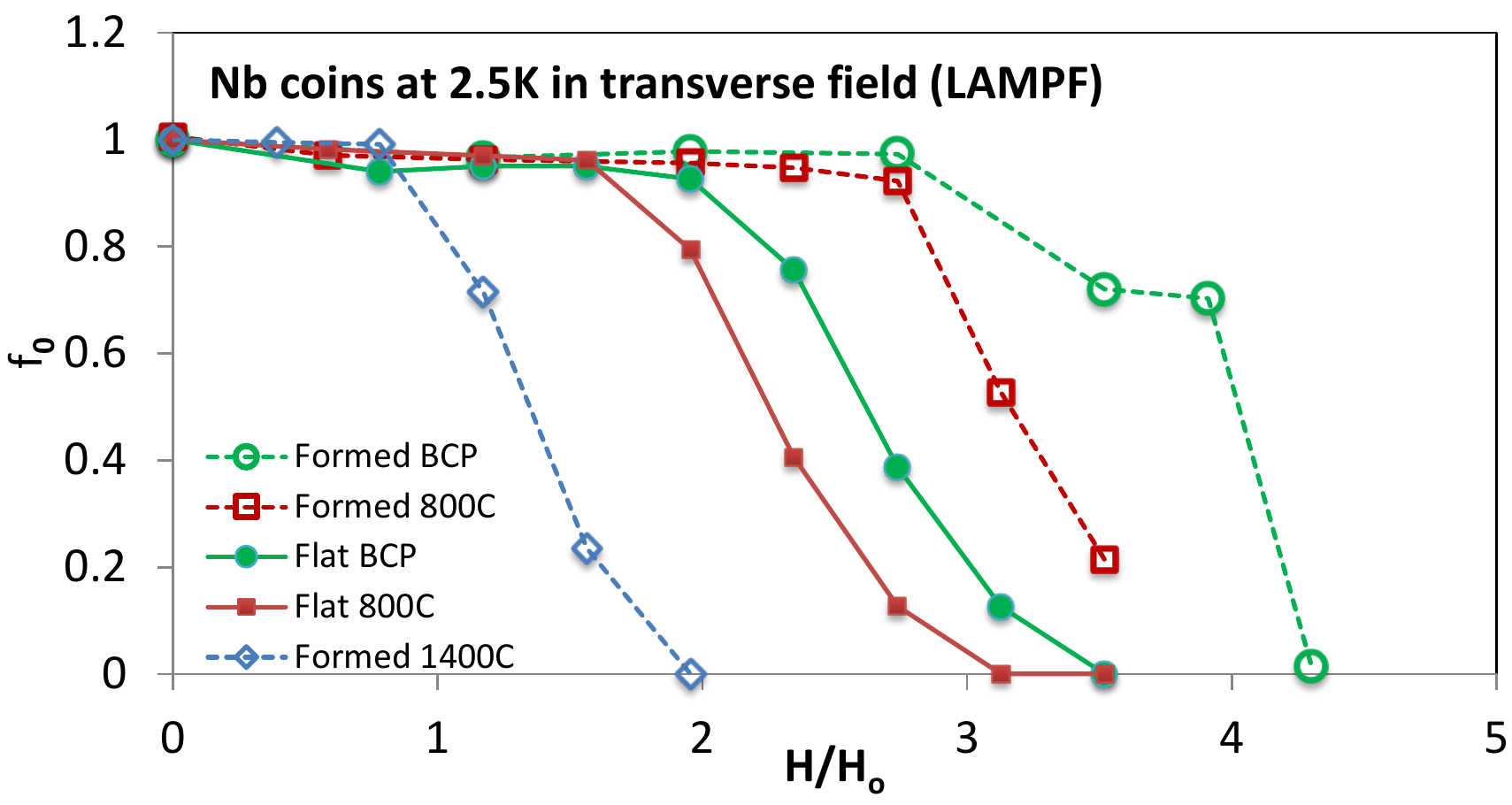}
    \caption{Fit parameter $f_0$ signifying the volume fraction probed by the muons which is in the field free Meissner state as a function of applied field for formed and flat geometries. Formed samples are denoted with dashed lines. Red lines are samples treated at \unit[800]{$\degree$C} for \unit[4]{hours}; green lines are for samples after BCP; blue line is a sample annealed at \unit[1400]{$\degree$C}. This plot shows that forming increases the pinning strength, which is virtually eliminated by a following annealing at \unit[1400]{$\degree$C}.}
    \label{fig:formed}
\end{figure}

In the next study the role of forming as a source of pinning is explored. Here the samples were cut using wire EDM from a \unit[1.3]{GHz} half cell (dumb-bell). The formed samples are treated with standard BCP and \unit[800]{$\degree$C} bake-out and are compared to flat samples with similar treatments. The results of the study are shown in \refF{fig:formed}. Pinning in formed samples delays flux entry to three times higher field as compared to the same sample after annealing at \unit[1400]{$\degree$C}. A \unit[800]{$\degree$C} treatment does relax pinning somewhat in both flat and formed cases. 


\subsection{Transverse ellipsoid results}
\begin{figure}[htbp]
    \centering
    \includegraphics[width=\columnwidth]{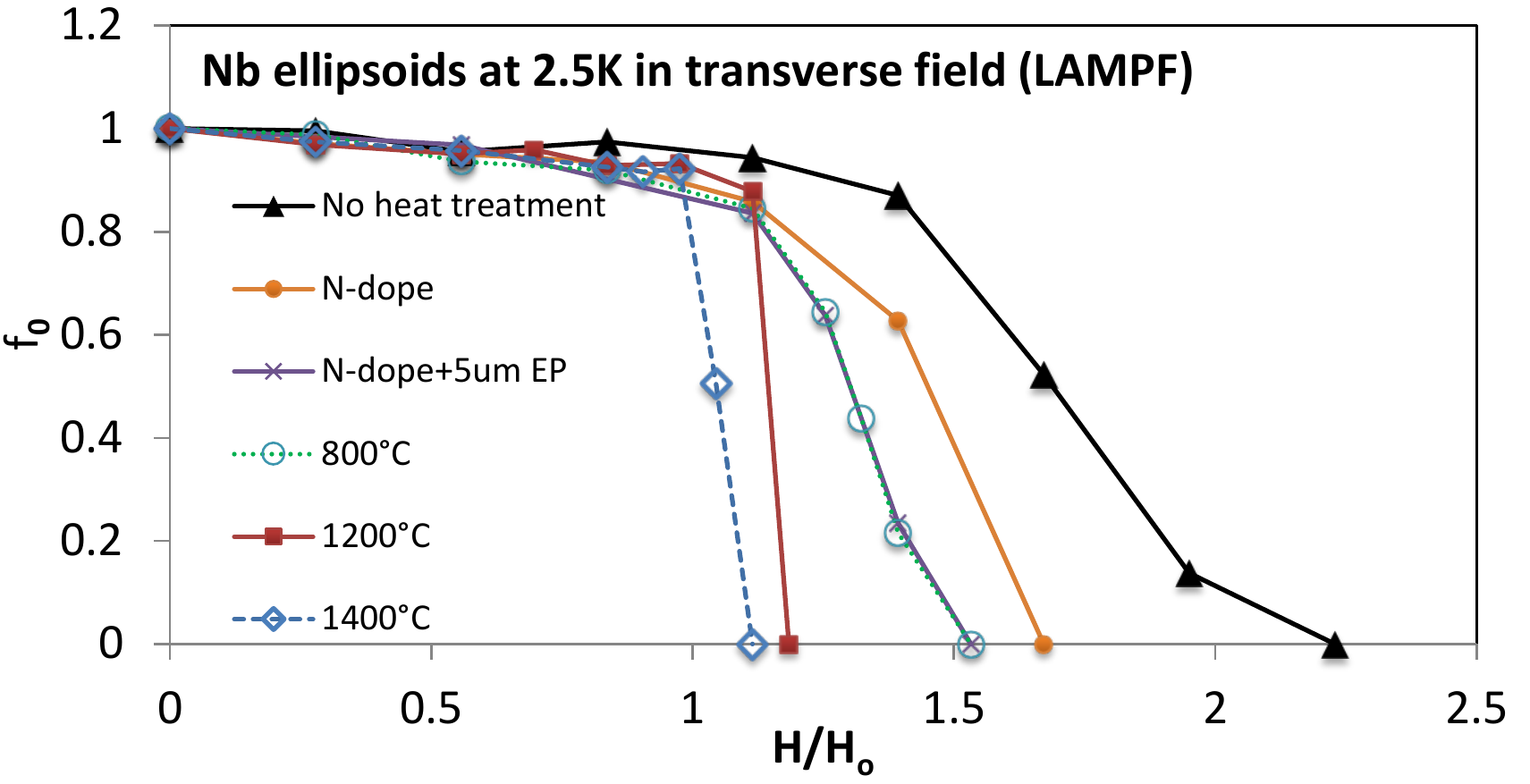}
    \caption{Fit parameter $f_0$ signifying the volume fraction probed by the muons which is in the field free Meissner state as a function of applied field for ellipsoid samples with different heat treatments measured in the LAMPF spectrometer. }
    \label{fig:Bullets_heat_treatments}
\end{figure}

The transverse ellipsoid is positioned as in \refF{fig:FieldConfiguration}(c) with the long axis coincident with the muon beam and aligned with the applied field. The muons are localized on the sample with a \unit[8]{mm} diameter mask identical to the transverse sample study. Here the mask confines the muons to the pole of the ellipsoid. The ellipsoids received various bulk and surface treatments as with the coins. In general, pinning is less predominant in the ellipsoid due to the lower demagnetization factor. However, pinning is still a factor as the fluxoids will nucleate at the equator and must overcome pinning to move to the pole. Heat treatments to \unit[1400]{$\degree$C} are shown to be effective to strongly reduce pinning as with the transverse coins. Figure \ref{fig:Bullets_heat_treatments} shows results from three ellipsoids with four hour heat treatments of \unit[1400]{$\degree$C}, \unit[1200]{$\degree$C}, and \unit[800]{$\degree$C} respectively. In both \unit[1400]{$\degree$C} and \unit[1200]{$\degree$C}, the field entry has a sharp threshold characteristic of uniform entry while the \unit[800]{$\degree$C} sample shows entry of fields at the pole for higher applied fields with an extended tail to $1.5 H_0$. 

Another set of studies was done comparing N-doped and non-N-doped material. The N-doping  was done at FNAL \cite{grassellino2013nitrogen}. The doping involves heating a sample to \unit[800]{$\degree$C}  for four hours and injecting N$_2$ gas near the end of the treatment. The results from these samples are also displayed in \refF{fig:Bullets_heat_treatments} where the characteristic flux entry is given for samples with EP (\unit[30]{$\mu$m}), with EP+N-dope, with EP+N-dope+EP (\unit[5]{$\mu$m}), and with EP+\unit[800]{$\degree$C} heat treatment. The results indicate that the \unit[800]{$\degree$C} bake (including N-dope) in all samples reduces the pinning compared to the sample with only EP. Further, the N-dope (a surface process) increases the pinning while the standard EP removal of \unit[5]{$\mu$m} reduces the pinning back to the same levels after only \unit[800]{$\degree$C} treatment. This will be examined in more detail in the following section.

\subsection{Parallel ellipsoid results}
\begin{figure}[htbp]
    \centering
    \includegraphics[width=\columnwidth]{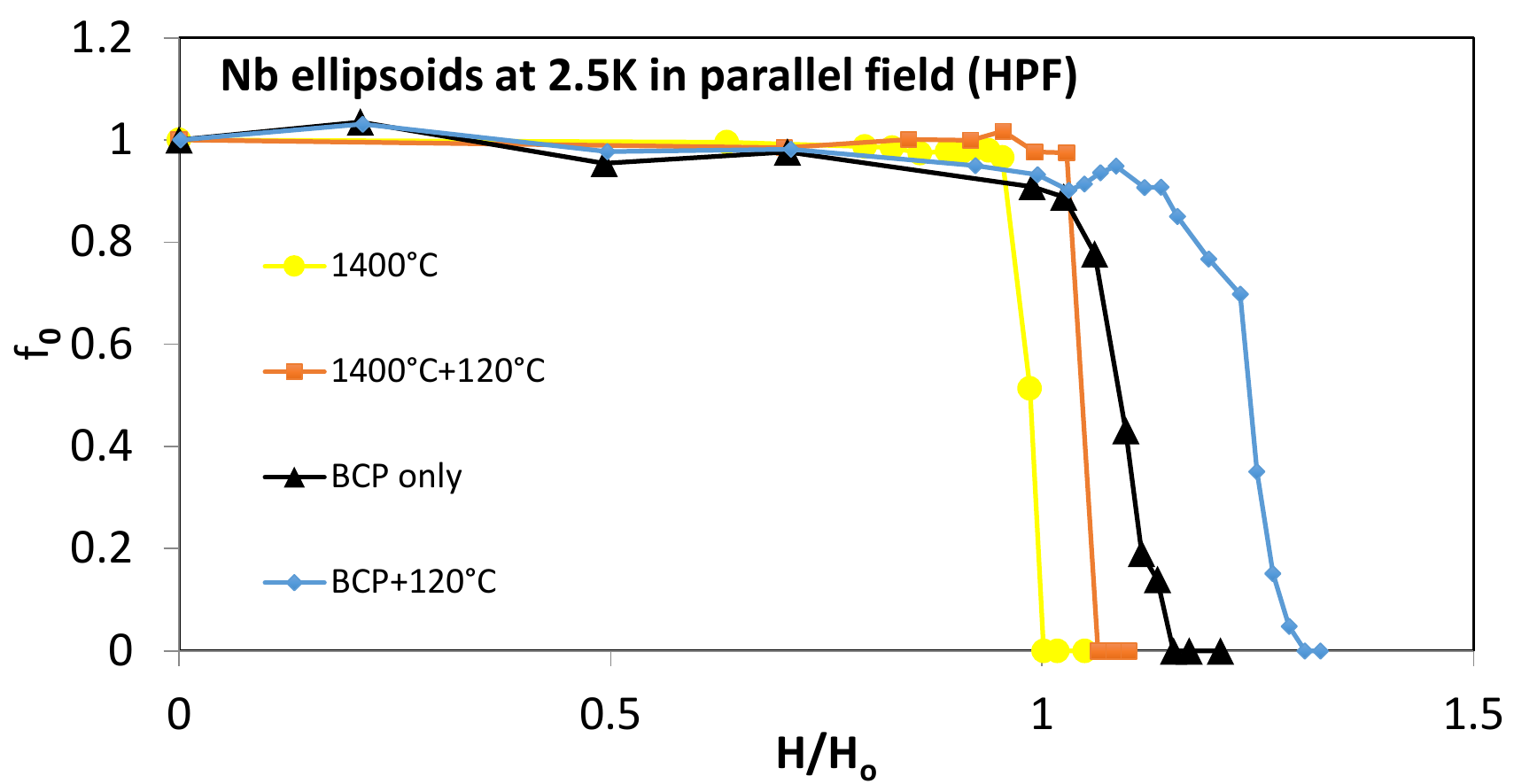}
    \caption{Fit parameter $f_0$ signifying the volume fraction probed by the muons which is in the field free Meissner state as a function of applied field for ellipsoid samples measured in the HPF spectrometer.}
    \label{fig:Smallbullets}
\end{figure}

The parallel ellipsoid geometry with the muons applied at the equator as displayed in \refF{fig:FieldConfiguration}(d) should be the least sensitive to pinning since the flux nucleates at the equator in the same location as the muons. The results of three samples tested in this geometry are displayed in \refF{fig:Smallbullets}. All samples were first etched (BCP) to remove \unit[100]{$\mu$m} material. One sample was then annealed at \unit[1400]{$\degree$C}. After the initial test it was baked at \unit[120]{$\degree$C} for \unit[48]{hours} together with one of the two samples which was not annealed. The results (\refF{fig:Smallbullets}) show that the field of first entry as sampled by the muon is dependent on the \unit[1400]{$\degree$C} anneal treatment and therefore suggest that the parallel ellipsoid geometry is also somewhat sensitive to pinning. Note that the muons are not stopped directly at the surface but about \unit[130]{$\mu$m} deep in the bulk. For the spot probed by the muons to remain in a field free state above $H\msub{nucleate}$ flux needs to be pinned in this \unit[130]{$\mu$m} layer. For comparison, the ellipsoid has a radius of \unit[6.3]{mm} at the equator. The hypothesis that a layer of a few $\mu$m can pin vortices is consistent with the finding that a nitrogen doped transverse ellipsoid showed stronger pinning compared to one which was only baked at the same temperature of \unit[800]{$\degree$C}, see \refF{fig:Bullets_heat_treatments}. Furthermore, after an additional electropolishing of only \unit[5]{$\mu$m}, the pinning strength of this sample was found to be identical to the one without doping.   

The low temperature baking increases $H\msub{a$\mid$entry}$ independent of whether the sample has been previously annealed. An increased $H\msub{entry}$ is consistent with an enhanced surface barrier preventing flux entry at $H\musb{c1}$ as recently proposed by Checchin et al. \cite{checchin2016ultimate}. Another explanation is increased surface pinning. Results with coins in transverse geometry have shown that baking at \unit[120]{$\degree$C} does not increase the bulk pinning strength\footnote{Also note that the edge boundary delaying flux entry make measurements in transverse coin not sensitive enough to measure small changes in $H\msub{entry}$.}. However, one can argue that the parallel geometry is more sensitive to surface pinning, because the flux lines can pin at the surface and delay migration to the muon implantation site, \unit[130]{$\mu$m} in the bulk. In the final discussion of the results we will revisit these two hypothesis, taking into account also the measurements on coated samples.

\subsection{Coated Samples}
This $\mu$SR method developed and commissioned using Nb samples has also been used to characterize Nb samples coated with higher $T\msub{c}$ materials. For example, collaborators at Cornell University have coated a standard Nb coin and ellipsoids for both spectrometers with a \unit[2]{$\mu$m} coating of Nb$_3$Sn using their standard recipe \cite{posen2015proof}. The same coin sample was tested in both transverse and parallel geometry. In total all four available arrangements (\refF{fig:FieldConfiguration}) have been used to test this coating. 

The muons are deposited at about \unit[130]{$\mu$m} into the Nb bulk. When the maximum field at the surface of the superconductor exceeds $H\msub{c1}$ of the material (or in case of a surface barrier, the superheating field $H\msub{sh}$) the material will enter the mixed phase. Depending on the temperature and applied field, Meissner screening could come either from the Nb$_3$Sn coating or the Nb bulk. Values of $H\msub{c1}(0K)$ and $H\msub{c}(0K)$ for niobium \cite{finnemore1966superconducting} of high purity and Nb$_3$Sn close to stoichiometry \cite{godeke2006nb3sn} are taken from the literature, while $H\msub{sh}$ is calculated from \cite{PhysRevB.83.094505}:
\begin{equation}
\frac{H\msub{sh}(\kappa)}{\sqrt{2}H\msub{c}}\approx\frac{\sqrt{10}}{6}+\frac{0.3852}{\kappa},
\end{equation}
where $\kappa$ is the Ginsburg-Landau parameter and $H\msub{c}$ the critical thermodynamic field. Table\,\ref{tab:MaterialParameters} lists $H\msub{c1}$ and $H\msub{sh}$ for both materials as well as the material parameters $\kappa$ and $T\msub{c}$ which are necessary to calculate $H\msub{sh}$ and the temperature dependence of the critical fields. For the following we assume that surface imperfections or geometry effects mean that $H\msub{c1}$ is the limitation for vortex penetration. We will revisit this assumption below. At \unit[2.5]{K} both the Nb and Nb$_3$Sn are superconducting and surface currents will be set up in the Nb$_3$Sn layer until $H\msub{c1}(T)$[Nb$_3$Sn] and in the Nb London layer from $H\msub{c1}$(T)\,[Nb$_3$Sn]$<H\msub{interface} <H\msub{c1}$(T)[Nb] with the Nb$_3$Sn coating in the vortex state, where $H\msub{interface}$ is the field at the interface between the Nb$_3$Sn layer and the Nb. 
\begin{table}
	\centering
		\begin{tabular}{l|c|c}
		Property & Nb & Nb$_3$Sn \\
		\hline
		$T\msub{c}$ & 9.25 & 18 \\
		$\mu_0 H\msub{c1}$(0K) [mT] & 174 & 38 \\
		$\mu_0 H\msub{c}$(0K) [mT] & 199 & 520 \\
		$\mu_0 H\msub{sh}$(0K) [mT] & 240 & 380 \\
		$\kappa$(0K) & 1.4 & 34 \\
		\end{tabular}
	\caption{Material parameters of Nb and Nb$_3$Sn.}
	\label{tab:MaterialParameters}
\end{table}

\begin{figure}[htbp]
    \centering
    \includegraphics[width=\columnwidth]{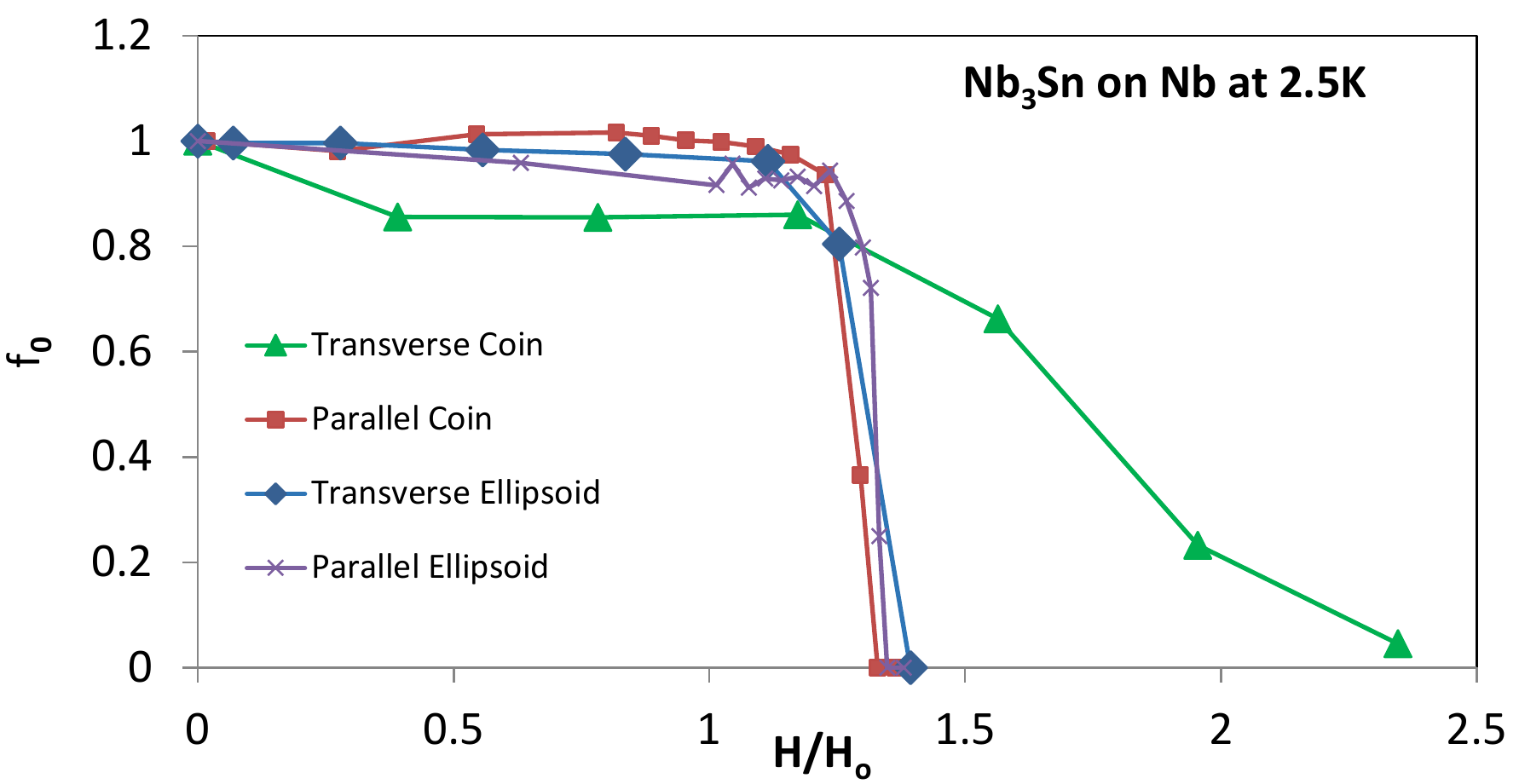}
    \caption{Fit parameter $f_0$ signifying the volume fraction probed by the muons which is in the field free Meissner stateas a function of applied field for Nb$_3$Sn coated Nb samples in four geometries. }
    \label{fig:Nb3Sn_geometries}
\end{figure}

\begin{figure}[htbp]
    \centering
    \includegraphics[width=\columnwidth]{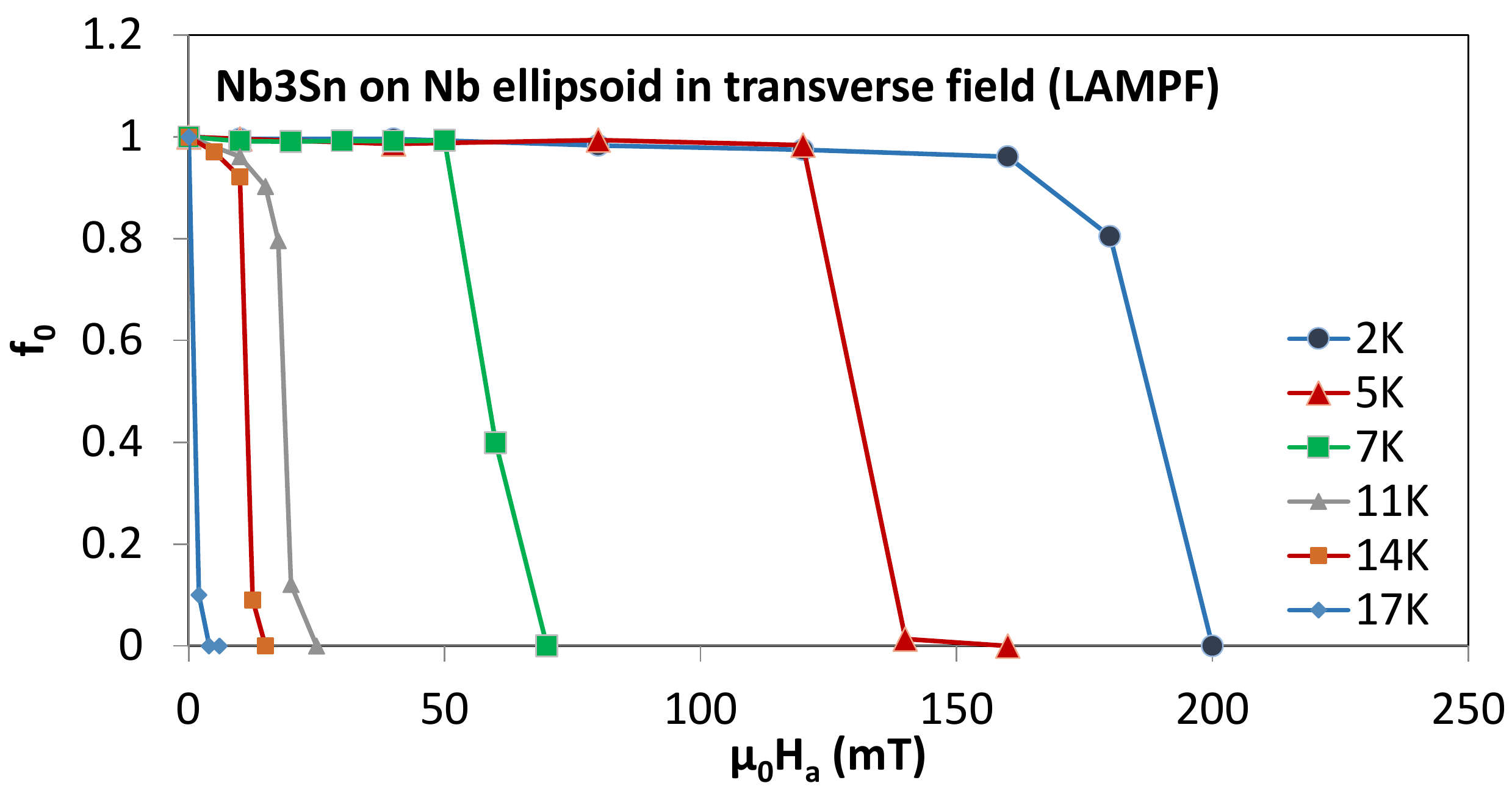}
    \caption{Fit parameter $f_0$ signifying the volume fraction probed by the muons which is in the field free Meissner state as a function of applied field for a Nb$_3$Sn coated Nb ellipsoid measured at different temperatures in the LAMPF spectrometer. }
    \label{fig:Nb3Sn_bullet}
\end{figure}

\begin{figure}[htbp]
    \centering
    \includegraphics[width=\columnwidth]{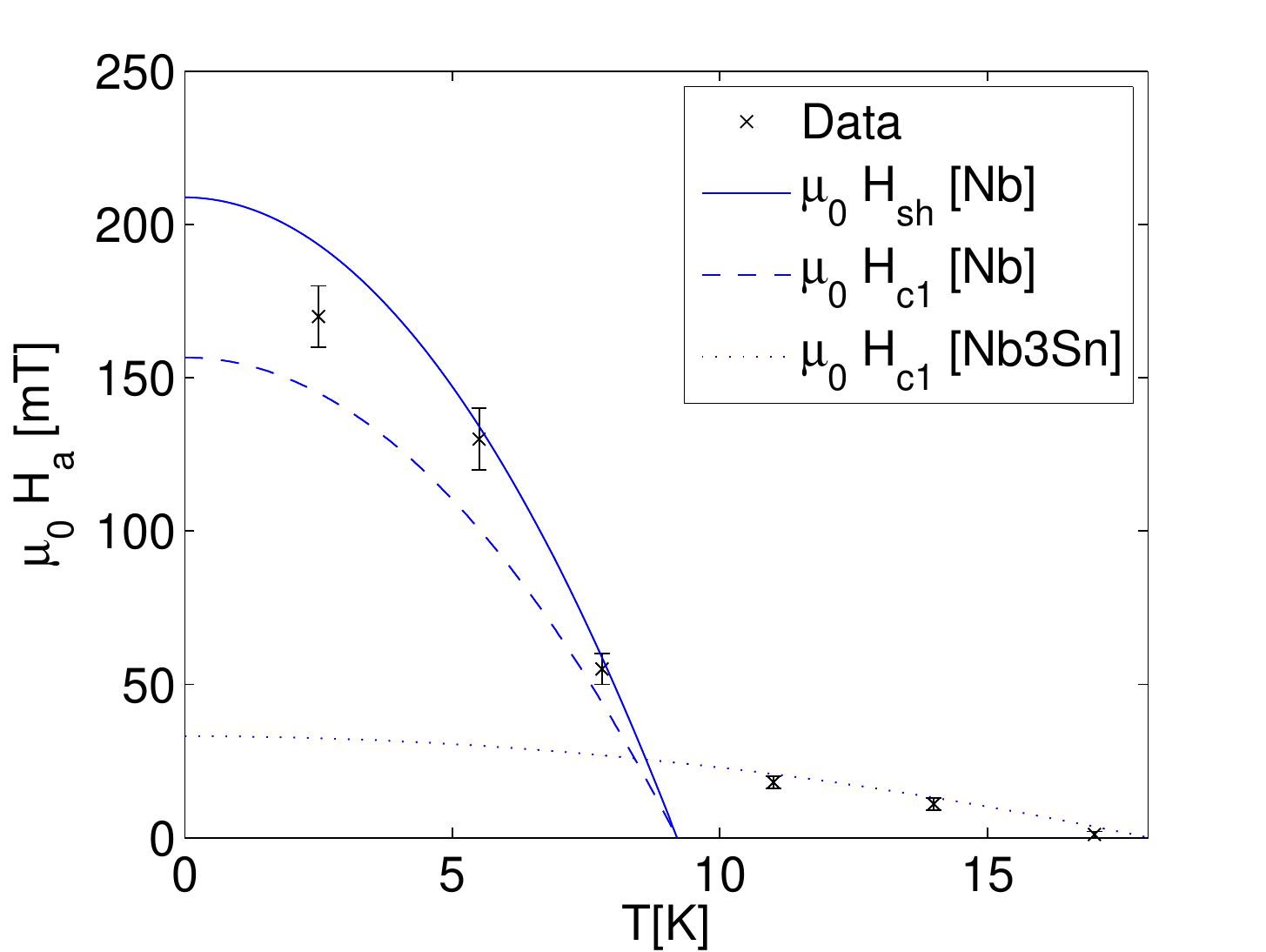}
    \caption{Measured field of first flux entry of a Nb$_3$Sn coated Nb ellipsoid. The lines are predictions for the superheating field $H\msub{sh}$ of Nb and the lower critical field $H\msub{c1}$ of Nb$_3$Sn and Nb taking into account the demagnetization factor of this geometry $N=0.13$.}
    \label{fig:Hsh}
\end{figure}

All four data sets measured at \unit[2.5]{K} are summarized in \refF{fig:Nb3Sn_geometries}. The data is normalized to $H_0$ as expected for Nb for convenient comparison with data presented above. 
The transverse coin results indicate that the bulk pinning in the sample is rather weak, when compared to the two cases shown in \refF{fig:UntreatedSamples} and \ref{fig:1400C}. This is understandable since the Nb$_3$Sn application involves a heat treatment to \unit[1100]{$\degree$C} \cite{posen2014advances}. The transverse ellipsoid results compare closely to the parallel results which is also indicative of low bulk pinning. In the transverse coin geometry, $f_0$ is reduced by about \unit[10]{\%} when a field is applied compared to its zero-field value. The reduction in field free area can be explained by noting that a \unit[1]{mm} hole is drilled in the coin center to allow coating in the furnace (\refF{fig:Samples}). Muons passing through this hole and getting stopped in the sample plate will sense the magnetic field, spin rotate and thus reduce $f_0$. From the combined results we find that the coating has pushed out the field of first flux entry to about \unit[1.3]{times} the standard Nb values, meaning that $H\msub{nucleate$\mid$0K}$ is enhanced to $\approx$\unit[240]{mT}, which is consistent with the superheating field $H\msub{sh}$ of niobium. The transverse ellipsoid sample has been measured at several temperatures above and below \unit[9.25]{K}, the critical temperature of niobium, see \refF{fig:Nb3Sn_bullet} and \,\ref{fig:Hsh}. For temperatures below \unit[9.25]{K}, the measured $H\msub{a$\mid$entry}$(T) is consistent with $H\msub{sh}$(T) of niobium. Above \unit[9.25]{K}, the data is consistent with flux entry at the lower critical field $H\msub{c1}$ of Nb$_3$Sn \cite{godeke2006nb3sn}. In \cite{posen2014advances} $H\msub{c1}$ was derived by extracting material parameters from Nb$_3$Sn SRF cavities prepared in the same furnace with the same coating parameters. These results are consistent with our measurements. This indicates that indeed $H\msub{c1}$[Nb$_3$Sn] is measured here. 

%
%



\section{Summary and Discussion}
The $\mu$SR technique applied to SRF materials has been extended by adding a dedicated spectrometer enabling measurements in strong parallel fields to the TRIUMF $\mu$SR facility and using different sample geometries. These various sample shapes and test configurations are crucial to the interpretation of the results. 

Bulk pinning in the material changes considerably depending on heat treatments. A \unit[1400]{$\degree$C} annealing virtually eliminates pinning and subsequent BCP does not erase this effect. For such annealed RRR niobium samples, measured in the HPF spectrometer with the field applied parallel to the sample surface, the field of first flux entry $H\msub{entry}$ is found consistent with literature values of $H\msub{c1}$. Generally, surface treatments such as \unit[120]{$\degree$C}, HF rinsing, and BCP of a few $\mu$m do not change the bulk pinning strength, showing that pinning is a bulk effect. Nitrogen doping however yields a slight increase in pinning strength which is erased by a subsequent \unit[5]{$\mu$m} BCP treatment. Effective pinning centers need to have a size on the order of the coherence length, which is in case of niobium is \unit[39]{nm}. It is therefore not surprising that \unit[120]{$\degree$C} baking and HF rinsing, affecting only a few nm of the surface, have no effect on the bulk pinning strength. In the case of N-doping without EP, the interpretation is that a dirty layer of microns depth can increase the near surface pinning and delay flux migrating into the ellipsoid where the muons are imbedded. 

Pinning is an important parameter for SRF applications. In order to achieve the lowest residual resistance, as required for CW applications, shielding of the earth's magnetic field alone is not sufficient but residual flux needs to be expelled \cite{posen2016efficient}. Several studies have directly addressed flux expulsion for different cavity treatments. However, there are only a few dedicated material studies using magnetometry to directly measure the pinning strength of SRF materials. Casalbuoni et al. have used cylinders cut from sheets \cite{casalbuoni2005surface}. The advantage of our method is that the muons are implanted locally allowing to distinguish better between geometrical edge pinning and intrinsic pinning from the material itself. Ashavai et al. avoid geometrical constraints by using long cylinders with very low demagnetization factors \cite{ashavai2012flux}. This method, unlike ours, does therefore not allow to use samples cut out from niobium sheet or from cavities characterized by RF and temperature mapping.


For pulsed applications of SRF technology, the maximum achievable accelerating gradient is the figure of merit. The intrinsic material parameter determining the maximum achievable accelerating gradient is the field of first flux entry, $H\musb{entry}$. For samples which have been baked at \unit[120]{$\degree$C} or coated with a Nb$_3$Sn overlayer we find $H\msub{entry}>H\msub{c1}$. For Nb$_3$Sn, $H\msub{entry}$ is pushed up to the superheating field of Nb, $H\msub{sh}$. If measured above $T\msub{c}[Nb]$, $H\msub{entry}$ is found consistent with $H\msub{c1}$ of Nb$_3$Sn. This result can be interpreted by an enhanced surface pinning or an increased intrinsic $H\musb{entry}$. The result of the \unit[120]{$\degree$C} baked samples is consistent with both hypotheses. The fact that for Nb$_3$Sn, if measured above $T\msub{c}[Nb]$=\unit[9.25]{K} $H\msub{entry}$ and its temperature dependence are found consistent with $H\msub{c1}[Nb_3Sn]$ show that the \unit[2]{$\mu$m} Nb$_3$Sn layer does not delay the penetration of the flux to the muon implantation site \unit[130]{nm} in the bulk. Furthermore, if this layer would provide pinning and delaying flux entry, for measurements below $T\msub{c}[Nb]$=\unit[9.25]{K}, one would not expect to find a temperature dependence consistent with pure niobium, but rather a relation depending on $T\msub{c}$ of Nb$_3$Sn as well. 
Furthermore, experiments presented elsewhere on annealed RRR niobium samples coated with MgB$_2$ layers between 50 and \unit[300]{nm} also find $H\msub{entry}(T)$ consistent with $H\msub{sh}(T)$[Nb] \cite{Junginger_superheating}. While these are strong arguments that the method developed here is well suited to probe the field of first flux entry into SRF materials, it cannot directly measure flux penetration in the London layer of a few nm. Such a method would have to be based on low energy muon spin rotation \cite{prokscha2008new} or $\beta$-NMR \cite{morris2013beta}.

\section{Conclusion}
A technique has been developed to measure the pinning strength and the field of first flux entry, $H\musb{entry}$, of SRF materials. If annealed substrates are used it is possible to measure $H\musb{entry}$ of layered structures and test theoretical predictions for multilayer structures proposed for next generation SRF cavities \cite{Gurevich_mulitlayers, Gurevich2015, kubo2014radio}. This has encouraged further investigations with different materials and layer thicknesses. The results from these studies are presented elsewhere \cite{Junginger_superheating}.   

\section{Acknowledgment}
This project has been supported by a Marie Curie International Outgoing Fellowship within the EU Seventh Framework Programme for Research and Technological Development (2007-2013).
Thanks to PAVAC for providing the formed cutouts, Anna Grassellino for providing the N-doping, Sam Posen, Matthias Liepe, and Daniel Leslie Hall for providing the Nb$_3$Sn treatments. Thanks to Gerald Morris, Bassam Hitti, Donald Arseneau, Deepak Vyas, Rahim Abasalti, and Iain McKenzie from the TRIUMF CMMS support team. Thanks to Bhalwinder Waraich for fabrication of the ellipsoid and flat coin samples. Thanks to the following people for participating to the experiment: D. Bazyl, D. Azzoni Gravel, Z. He, Y. Ma, L. Yang, Z. Yao, and H. Zhang.



\end{document}